\documentclass[a4paper,11pt]{article}
\pdfoutput=1 

\usepackage{jheppub} 

\usepackage[T1]{fontenc} 

\usepackage{bbm,latexsym,amssymb}
\usepackage{color}

\def\be{\begin{equation}}
\def\ee{\end{equation}}
\def\ba{\begin{eqnarray}}
\def\ea{\end{eqnarray}}
\def\ra{\rightarrow}
\def\vb#1{\vbox to #1 pt{}}

\title{\boldmath $h \rightarrow Z \gamma$
in the complex two Higgs doublet model}

\author[a]{Duarte Fontes,}
\author[a]{J. C. Rom\~{a}o}
\author[b,a,1]{and Jo\~{a}o P. Silva\note{Corresponding author.}}

\affiliation[a]{Centro de F\'{\i}sica Te\'{o}rica de Part\'{\i}culas (CFTP),
    Instituto Superior T\'{e}cnico, Universidade de Lisboa,
    1049-001 Lisboa, Portugal}
\affiliation[b]{Instituto Superior de Engenharia de Lisboa - ISEL,
	1959-007 Lisboa, Portugal}

\emailAdd{duartefontes@tecnico.ulisboa}
\emailAdd{jorge.romao@tecnico.ulisboa}
\emailAdd{jpsilva@cftp.ist.utl.pt}

\abstract{The latest LHC data confirmed the existence of a
Higgs-like particle and made interesting measurements on its decays
into $\gamma \gamma$, $Z Z^\ast$, $W W^\ast$, $\tau^+ \tau^-$,
and $b \bar{b}$.
It is expected that a decay into $Z \gamma$ might be measured
at the next LHC round, for which there already exists an upper bound.
The Higgs-like particle could be a mixture of scalar with a
relatively large component of pseudoscalar.
We compute the decay of such a mixed state into $Z \gamma$,
and we study its properties in the context of the complex two
Higgs doublet model,
analysing the effect of the current measurements on
the four versions of this model.
We show that a measurement of the $h \rightarrow Z \gamma$ rate at
a level consistent with the SM can be used to place interesting
constraints on the pseudoscalar component.
We also comment on the issue of a wrong sign Yukawa coupling
for the bottom in Type II models.}

\begin{document}
\maketitle
\flushbottom

\section{Introduction}
\label{sec:intro}

The ATLAS \cite{atlas} and CMS  \cite{cms} experiments at LHC
have detected a particle with properties closely resembling
those of the SM Higgs,
in the decay channels $\gamma \gamma$, $Z Z^\ast$, $W W^\ast$,
and $\tau^+ \tau^-$,
with errors of order $20\%$.
Decays into $b \bar{b}$ are only detected at LHC and the
Tevatron in connection with
the associated $Vh$ production mechanism,
with errors of order $50\%$
\cite{cms:bb, Tuchming:2014fza}.
Up-to-date LHC results can be found in refs.~\cite{atlas2, cms2}.

The discovery of $pp \ra h \ra \gamma \gamma$ can be seen
as the poster child of quantum field theory:
the dominating production through gluon-gluon fusion occurs at one loop;
and so does the decay into $\gamma \gamma$.
Recently,
ATLAS \cite{atlas:Zph} and CMS \cite{cms:Zph} have reported on
the search for another loop-decay, $h \ra Z \gamma$,
finding upper bounds of order ten times the SM expectation
at the $95\%$ confidence level.
This is expected to be the next interesting channel to be
measured in the upcoming LHC run.

As the newfound particle is further probed,
there are two interesting questions that
will be considered: i) is the new particle
purely scalar, or does it have some
pseudoscalar component?; ii) how many scalars
are there?
On the first issue,
we know from the existence of $h \ra V V$
that $h$ cannot be purely pseudoscalar (henceforth, $V = W, Z$).
There are some experimental bounds on the likelihood that
the 125 GeV particle is a pure pseudoscalar
\cite{Chatrchyan:2012jja, GABRIELLI:2014sra},
but we are interested here in the possibility that the 125 GeV
state is a mixture of scalar and pseudo-scalar components.
On the second issue,
although there have been some experimental fluctuations,
there is currently no sign of another scalar.
However,
the limits are rather loose and the possibility remains that
there are further scalar, including charged,
that have evaded detection because its couplings are
not too large.
For example,
in two Higgs doublet models,
the fact that the observed scalar has couplings to two vector bosons
in line with SM expectations forces the couplings of the
heavier scalar to two vector bosons to be small.

The main objectives of this article are two-pronged.
Firstly,
we discuss the production and decays of a spin zero state which is a mixture
of scalar and pseudoscalar, with special emphasis on a detailed
discussion $h \ra Z \gamma$.
The details are contained in the appendices.
Secondly,
we analyze the current bounds on the complex two Higgs doublet
model (C2HDM), where the lightest Higgs is in general a mixture
of scalar and pseudoscalar.

The article is organized as follows.
In section~\ref{sec:C2HDM} we summarize the C2HDM and introduce our
notation. In section~\ref{sec:simulation} we discuss,
in turn,
current constraints and future reach on the four types
of flavour couplings (Type I, Type II, Lepton Specific, and Flipped).
As far as we know,
this is the first update on the first two types,
and the first discussion
of the Lepton Specific and Flipped models to use the latest Run 1
data from LHC.
In particular,
we also discuss the effect of future experiments,
and
what might be learned from $h \ra Z \gamma$ at
LHC's Run 2.
In section~\ref{sec:wrong_sign},
we discuss the possibility that the scalar component of
the $h_1 b \bar{b}$ coupling has a sign opposite to the SM.
We relate this with the situation in the real 2HDM,
which has received recent interest.
Finally, we draw our conclusions in section~\ref{sec:conclusions}.

For completeness we collect in the appendices all the
expressions needed for the production and decay of a
Higgs boson which has a mixture of scalar and pseudo-scalar
components;
this includes the neutral scalars of the
most general 2HDM.
In particular,
the expressions for the one loop decays
are given in a form that can be useful for other models with a more
general Higgs boson sector than the SM.
We also compare our results
with those that can be found in the literature.

\section{The complex two Higgs doublet model}
\label{sec:C2HDM}

We consider a model with two Higgs doublets, $\phi_1$ and $\phi_2$,
with the $Z_2$ symmetry $\phi_1 \ra \phi_1, \phi_2 \ra -\phi_2$
violated softly.
The Higgs potential can be written as \cite{ourreview}
\ba
V_H
&=&
m_{11}^2 |\phi_1|^2
+ m_{22}^2 |\phi_2|^2
- m_{12}^2\, \phi_1^\dagger \phi_2
- (m_{12}^2)^\ast\, \phi_2^\dagger \phi_1
\nonumber\\*[2mm]
&&
+\, \frac{\lambda_1}{2} |\phi_1|^4
+ \frac{\lambda_2}{2} |\phi_2|^4
+ \lambda_3 |\phi_1|^2 |\phi_2|^2
+ \lambda_4\, (\phi_1^\dagger \phi_2)\, (\phi_2^\dagger \phi_1)
\nonumber\\*[2mm]
&&
+\, \frac{\lambda_5}{2} (\phi_1^\dagger \phi_2)^2
+ \frac{\lambda_5^\ast}{2} (\phi_2^\dagger \phi_1)^2.
\label{VH}
\ea
Hermiticity implies that all couplings are real,
except $m_{12}^2$ and $\lambda_5$.
If $\textrm{arg}(\lambda_5) \neq 2\, \textrm{arg}(m_{12}^2)$,
then the phases cannot be removed.
This is known as the complex two Higgs doublet model
(C2HDM),
and has been studies extensively in\footnote{Our
notation differs from theirs, and agrees with \cite{ourreview},
in that $2 m_{11}^2 = - m_{11}^2(\textrm{theirs})$,
$2 m_{22}^2 = - m_{22}^2(\textrm{theirs})$,
and
$2 m_{12}^2 = m_{12}^2(\textrm{theirs})$.}
refs.~\cite{Ginzburg:2002wt, Khater:2003wq,
ElKaffas:2007rq, El Kaffas:2006nt, WahabElKaffas:2007xd,
Osland:2008aw, Grzadkowski:2009iz, Arhrib:2010ju, pseudo}.
If $\textrm{arg}(\lambda_5) = 2\, \textrm{arg}(m_{12}^2)$,
then we can choose a basis where $m_{12}^2$ and $\lambda_5$ become real and,
if the vacuum expectation values (vev) of $\phi_1$ and $\phi_2$
are also real,
we talk about the real 2HDM.
Henceforth, it is implicit that the C2HDM and the real
2HDM have a softly broken $Z_2$ symmetry.

With a suitable basis choice,
we can take the vevs real:
\be
\langle \phi_1 \rangle = v_1/\sqrt{2},
\hspace{5ex}
\langle \phi_2 \rangle = v_2/\sqrt{2},
\label{v1v2}
\ee
and write the scalar doublets as
\be
\phi_1 =
\left(
\begin{array}{c}
\varphi_1^+\\
\tfrac{1}{\sqrt{2}} (v_1 + \eta_1 + i \chi_1)
\end{array}
\right),
\hspace{5ex}
\phi_2 =
\left(
\begin{array}{c}
\varphi_2^+\\
\tfrac{1}{\sqrt{2}} (v_2 + \eta_2 + i \chi_2)
\end{array}
\right).
\ee
With this convention,
$v = \sqrt{v_1^2 + v_2^2} = (\sqrt{2} G_\mu)^{-1/2} = 246$ GeV,
and the stationarity conditions become
\ba
-2\, m_{11}^2 &=&
- \textrm{Re}\left(m_{12}^2 \right) \frac{v_2}{v_1}
+ \lambda_1\, v_1^2 + \lambda_{345}\, v_2^2,
\nonumber\\
-2\, m_{22}^2 &=&
- \textrm{Re}\left(m_{12}^2 \right) \frac{v_1}{v_2}
+ \lambda_2\, v_2^2 + \lambda_{345}\, v_1^2,
\nonumber\\
2\, \textrm{Im}\left(m_{12}^2 \right) &=& v_1 v_2\, \textrm{Im}\left(\lambda_5 \right),
\label{stat_cond}
\ea
where $\lambda_{345}= \lambda_3 +  \lambda_4 + \textrm{Re}\left(\lambda_5 \right)$.

We can now transform the fields into the Higgs basis by
\cite{LS,BS}
\be
\left(
\begin{array}{c}
H_1\\
H_2
\end{array}
\right)
=
\left(
\begin{array}{cc}
c_{\beta} & s_{\beta}\\
- s_{\beta} & c_{\beta}
\end{array}
\right)
\left(
\begin{array}{c}
\phi_1\\
\phi_2
\end{array}
\right),
\ee
where $\tan{\beta} = v_2/v_1$,
$c_\beta = \cos{\beta}$, and $s_\beta = \sin{\beta}$.
The Higgs basis was introduced \cite{LS,BS} such that the
second Higgs does not get a vev:
\be
H_1 =
\left(
\begin{array}{c}
G^+\\
\tfrac{1}{\sqrt{2}} (v + H^0 + i G^0)
\end{array}
\right),
\hspace{5ex}
H_2 =
\left(
\begin{array}{c}
H^+\\
\tfrac{1}{\sqrt{2}} (R_2 + i I_2)
\end{array}
\right).
\ee
In this basis,
$G^+$ and $G^0$ are massless and,
in the unitary gauge,
will become the longitudinal components
of $W^+$ and $Z^0$,
respectively.
There remains a charged pair $H^\pm$ with mass $m_{H^\pm}$.

In the usual notation for the C2HDM,
$\eta_3 = I_2$, and the three neutral components
mix into the neutral mass eigenstates through
\be
\left(
\begin{array}{c}
h_1\\
h_2\\
h_3
\end{array}
\right)
= R
\left(
\begin{array}{c}
\eta_1\\
\eta_2\\
\eta_3
\end{array}
\right).
\label{h_as_eta}
\ee
The orthogonal matrix $R$ diagonalizes the
neutral mass matrix
\be
\left( {\cal M}^2 \right)_{ij} =
\frac{\partial^2 V_H}{\partial \eta_i\, \partial \eta_j},
\ee
through
\be
R\, {\cal M}^2\, R^T = \textrm{diag} \left(m_1^2, m_2^2, m_3^2 \right),
\ee
where $m_1 \leq m_2 \leq m_3$ are the masses of the neutral Higgs particles.
The matrix $R$ can be parametrized as \cite{ElKaffas:2007rq}
\be
R =
\left(
\begin{array}{ccc}
c_1 c_2 & s_1 c_2 & s_2\\
-(c_1 s_2 s_3 + s_1 c_3) & c_1 c_3 - s_1 s_2 s_3  & c_2 s_3\\
- c_1 s_2 c_3 + s_1 s_3 & -(c_1 s_3 + s_1 s_2 c_3) & c_2 c_3
\end{array}
\right)
\label{matrixR}
\ee
where $s_i = \sin{\alpha_i}$ and
$c_i = \cos{\alpha_i}$ ($i = 1, 2, 3$).
Without loss of generality,
the angles may be restricted to \cite{ElKaffas:2007rq}
\be
- \pi/2 < \alpha_1 \leq \pi/2,
\hspace{5ex}
- \pi/2 < \alpha_2 \leq \pi/2,
\hspace{5ex}
0 \leq \alpha_3 \leq \pi/2.
\label{range_alpha}
\ee

The relation between the Higgs basis and the mass basis is
\be
\left(
\begin{array}{c}
\eta_1\\
\eta_2\\
\eta_3
\end{array}
\right)
=
R_H
\left(
\begin{array}{c}
H^0\\
R_2\\
I_2
\end{array}
\right)
=
\left(
\begin{array}{ccc}
c_{\beta} & - s_{\beta} & 0\\
s_{\beta} & c_{\beta} & 0\\
0 & 0 & 1
\end{array}
\right)
\left(
\begin{array}{c}
H^0\\
R_2\\
I_2
\end{array}
\right).
\ee
Thus
\be
\left(
\begin{array}{c}
h_1\\
h_2\\
h_3
\end{array}
\right)
=
R
\left(
\begin{array}{c}
\eta_1\\
\eta_2\\
\eta_3
\end{array}
\right)
=
R\, R_H
\left(
\begin{array}{c}
H^0\\
R_2\\
I_2
\end{array}
\right).
\ee
The computation of the bounds from the oblique radiative corrections
in eqs.~(388) and (393) of ref.~\cite{ourreview} requires the matrix
$T = R_H^T R^T$ in eq.~(381) of ref.~\cite{ourreview}.

Given an arbitrary relative phase,
the Higgs potential in eq.~\eqref{VH} has 9 independent
parameters.
We follow ref.~\cite{El Kaffas:2006nt},
and trade these for $v$ and for the 8 input
parameters $\beta$, $m_{H^\pm}$,
$\alpha_1$, $\alpha_2$, $\alpha_3$,
$m_1$, $m_2$, and $\textrm{Re}(m_{12}^2)$.
With this choice,
$m_3$ is given by
\be
m_3^2 = \frac{m_1^2\, R_{13} (R_{12} \tan{\beta} - R_{11})
+ m_2^2\ R_{23} (R_{22} \tan{\beta} - R_{21})}{R_{33} (R_{31} - R_{32} \tan{\beta})}.
\label{m3_derived}
\ee
Of course,
we are only interested in those cases where $m_3^2 >0$,
and, due to our mass ordering,
$m_3^2 > m_2^2 > m_1^2$.
This places constraints on the relevant
parameter space.

The Higgs potential in eq.~\eqref{VH} can be reconstructed
through
\ba
v^2\, \lambda_1 &=&
- \frac{1}{\cos^2{\beta}}
\left[- m_1^2\, c_1^2 c_2^2 - m_2^2 (c_3 s_1 + c_1 s_2 s_3)^2
- m_3^2\, (c_1 c_3 s_2 - s_1 s_3)^2 + \mu^2\, \sin^2{\beta}
\right],
\nonumber\\*[2mm]
v^2\, \lambda_2 &=&
- \frac{1}{\sin^2{\beta}}
 \left[
- m_1^2\, s_1^2 c_2^2 - m_2^2\, (c_1 c_3 - s_1 s_2 s_3)^2
- m_3^2\, (c_3 s_1 s_2 + c_1 s_3)^2  + \mu^2\, \cos^2{\beta}
\right],
\nonumber\\*[2mm]
v^2\, \lambda_3 &=&
\frac{1}{\sin{\beta} \cos{\beta}}
\left[
\left(
m_1^2\, c_2^2
+ m_2^2\, (s_2^2 s_3^2 - c_3^2)
+ m_3^2\, (s_2^2 c_3^2 - s_3^2)
\right) c_1 s_1
\right.
\nonumber\\*[2mm]
& &
\hspace{10ex}
\left.
+\,
(m_3^2 - m_2^2) (c_1^2 - s_1^2) s_2 c_3 s_3
\right]
- \mu^2 + 2 m_{H^\pm}^2\, ,
\nonumber\\*[2mm]
v^2\, \lambda_4 &=&
m_1^2\, s_2^2 + ( m_2^2\,  s_3^2 + m_3^2\, c_3^2) c_2^2
+ \mu^2 - 2 m_{H^\pm}^2,
\nonumber\\*[2mm]
v^2\, \textrm{Re}(\lambda_5)
&=&
- m_1^2\, s_2^2 - (m_2^2\, s_3^2 + m_3^2\, c_3^2) c_2^2 + \mu^2,
\nonumber\\*[2mm]
v^2\, \textrm{Im}(\lambda_5)
&=&
\frac{2}{\sin{\beta}}
c_2
\left[
(- m_1^2 + m_2^2\, s_3^2 + m_3^2\, c_3^2) c_1 s_2
+ (m_2^2 - m_3^2) s_1 s_3 c_3
\right],
\label{reconstruct_lambdas}
\ea
where
\be
\mu^2 =
\frac{v^2}{v_1\, v_2}\, \textrm{Re}(m_{12}^2).
\label{eq:mu}
\ee
We have checked that,
using eq.~\eqref{m3_derived},
we reproduce the results in eq.~(B.1) of
ref.~\cite{Arhrib:2010ju}.

To compute the decays of the lightest Higgs we need the couplings
$h_1 V V$ ($V=W,Z$), $h_1 H^+ H^-$, and $h_1 \bar{f} f$ for
some fermion $f$.
These can be obtained from the Higgs potential,
the covariant derivatives,
and the Yukawa potential,
respectively.
As shown in ref.~\cite{pseudo},
the $h_1 V V$ and $h_1 H^+ H^-$
can be written,
respectively,
as in eqs.~\eqref{LVV} and
\eqref{LhHpHm},
with
\be
C = c_\beta R_{11} + s_\beta R_{12}
=
\cos{(\alpha_2)}\, \cos{(\alpha_1 - \beta)},
\label{C}
\ee
and
\be
- \lambda
=
c_\beta \left[ s_\beta^2 \lambda_{145} + c_\beta^2 \lambda_3 \right] R_{11}
+
s_\beta \left[ c_\beta^2 \lambda_{245} + s_\beta^2 \lambda_3 \right] R_{12}
+
s_\beta c_\beta\, \textrm{Im}(\lambda_5)\, R_{13},
\label{lambda_hHpHm}
\ee
where $\lambda_{145} = \lambda_1 - \lambda_4 - \textrm{Re}(\lambda_5)$
and $\lambda_{245} = \lambda_2 - \lambda_4 - \textrm{Re}(\lambda_5)$.
In order to preclude flavour changing interactions with the neutral Higgs,
each fermion sector must couple to only one Higgs.
In the usual notation,
up-type quarks couple to $\phi_2$,
so there are four possibilities according to the couplings
of down-type quarks and charged leptons.
In Type I (Type II) both couple to $\phi_2$ ($\phi_1$).
In Lepton Specific (Flipped),
down-type quarks couple to $\phi_2$ ($\phi_1$),
while charged leptons couple to $\phi_1$ ($\phi_2$).
The result can be written as in eq.~\eqref{LY},
with the coefficients $a + i b \gamma_5$ given
in table~\ref{tab:1}.
%
%
\begin{table}[tbp]
\centering
\begin{tabular}{|lcccccccc|}
\hline
 & & Type I  & & Type II & & Lepton & & Flipped \\
 & & & & & & Specific & & \\
\hline
Up  & &
$\tfrac{R_{12}}{s_{\beta}} - i c_\beta \tfrac{R_{13}}{s_{\beta}} \gamma_5$   & &
$\tfrac{R_{12}}{s_{\beta}} - i c_\beta \tfrac{R_{13}}{s_{\beta}} \gamma_5$  & &
$\tfrac{R_{12}}{s_{\beta}} - i c_\beta \tfrac{R_{13}}{s_{\beta}} \gamma_5$   & &
$\tfrac{R_{12}}{s_{\beta}} - i c_\beta \tfrac{R_{13}}{s_{\beta}} \gamma_5$  \\*[2mm]
Down  & &
$\tfrac{R_{12}}{s_{\beta}} + i c_\beta \tfrac{R_{13}}{s_{\beta}} \gamma_5$   & &
$\tfrac{R_{11}}{c_{\beta}} - i s_\beta \tfrac{R_{13}}{c_{\beta}} \gamma_5$    & &
$\tfrac{R_{12}}{s_{\beta}} + i c_\beta \tfrac{R_{13}}{s_{\beta}} \gamma_5$   & &
$\tfrac{R_{11}}{c_{\beta}} - i s_\beta \tfrac{R_{13}}{c_{\beta}} \gamma_5$    \\*[2mm]
Leptons  & &
$\tfrac{R_{12}}{s_{\beta}} + i c_\beta \tfrac{R_{13}}{s_{\beta}} \gamma_5$   & &
$\tfrac{R_{11}}{c_{\beta}} - i s_\beta \tfrac{R_{13}}{c_{\beta}} \gamma_5$    & &
$\tfrac{R_{11}}{c_{\beta}} - i s_\beta \tfrac{R_{13}}{c_{\beta}} \gamma_5$   & &
$\tfrac{R_{12}}{s_{\beta}} + i c_\beta \tfrac{R_{13}}{s_{\beta}} \gamma_5$   \\*[2mm]
\hline
\end{tabular}
\caption{\label{tab:1} Couplings of the fermions to the lightest scalar, $h_1$,
in the form $a + i b\gamma_5$ of eq.~\eqref{LY}.}
\end{table}

Looking back at eqs.~\eqref{h_as_eta} and \eqref{matrixR},
we realize that $|s_2|$ measures the pseudoscalar component
of the lightest neutral scalar, $h_1$.
Indeed,
when $s_2=0$, the pseudoscalar $\eta_3$ does not contribute
to $h_1$,
while,
when $c_2=0$ only the pseudoscalar $\eta_3$ contributes
to $h_1$.
That is,
\ba
|s_2| = 0\ \
& \Longrightarrow &
\ \ h_1\ \textrm{is a pure scalar},
\label{pure_scalar}
\\
|s_2| = 1\ \
& \Longrightarrow &
\ \ h_1\ \textrm{is a pure pseudoscalar}.
\label{pure_pseudoscalar}
\ea
This is confirmed by the form of the various couplings.
In fact,
the $h_1 V V$ coupling $C$ in eq.~\eqref{C} vanishes when $|s_2|=1$,
in agreement with the absence of a pseudoscalar coupling
with a vector boson/anti-boson pair.
Similarly,
when $|s_2|=1$ the only term in
$\lambda$ which survives is the term proportional
to $\textrm{Im}(\lambda_5)$ in eq.~\eqref{lambda_hHpHm}.
This is consistent with the fact that a pseudoscalar
can only couple to $H^+ H^-$ if there is explicit
CP violation in the Higgs potential.
Finally,
when $s_2=0$,
all $b$ coefficients in table~\ref{tab:1} (multiplying $i \gamma_5$)
vanish,
and $h_1$ couples to fermions as a pure scalar.
Similarly,
when $|s_2|=1$,
all $a$ coefficients in table~\ref{tab:1} vanish,
and $h_1$ couples to fermions as a pure pseudoscalar.

\section{Simulation procedure and results}
\label{sec:simulation}

For our fit procedure, we generate points in parameter space
with $m_1 = 125$ GeV,
the angles $\alpha_{1,2,3}$ within the intervals
of eq.~\eqref{range_alpha},
$1 \leq \tan{\beta} \leq 30$,
$ m_1 \leq m_2 \leq 900\, \textrm{GeV}$,
$-(900\, \textrm{GeV})^2 \leq m_{12}^2 \leq (900\, \textrm{GeV})^2$,
and $340\, \textrm{GeV} \leq m_{H^\pm} \leq 900\, \textrm{GeV}$
(Type II and Flipped),
or $100\, \textrm{GeV} \leq m_{H^\pm} \leq 900\, \textrm{GeV}$
(Type I and Lepton Specific).

The ranges for $m_{H^\pm}$ and $\tan{\beta}$ where chosen
to comply with the constraints from $Z \ra b \bar{b}$,
$b \ra s \gamma$, and other $B$-Physics results.
The constraints are basically the same in the complex and real
2HDM because the charged Higgs couplings to fermions coincide
-- see, for example, appendix C of \cite{Arhrib:2010ju}.
In Type II and Flipped,
$Z \ra b \bar{b}$ implies $\tan{\beta} \gtrsim 1$
while $b \ra s \gamma$ excludes values of $m_{H^\pm}$
below $360$ GeV, at the 95\% confidence level,
with only a very mild dependence on $\tan{\beta}$
\cite{Haber:1999zh,gfitter1,hermann,mahmoudi}.
In Type I and Lepton Specific,
$\tan{\beta} \gtrsim 1$ still holds,
but $m_{H^\pm}$ can be as low as $\sim 90 \textrm{GeV}$,
even after the LHC results on
$pp \ra t \bar{t}$ with decay into $H^+ \bar{b}$
\cite{Aad:2012tj,Chatrchyan:2012vca}.
The ranges we have chosen for $m_{H^\pm}$
and $\tan{\beta}$ conform to rather conservative
bounds from these and other B-Physics experiments,
and, for comparison purposes,
were taken to coincide with the constraints
in refs.~\cite{Ferreira:2014naa,Fontes:2014tga},
in the CP conservative limit.

Given a set of input parameters,
$m_3^2$ is obtained from eq.~\eqref{m3_derived}.
With our conventions,
one should only take points where $m_3^2 > m_2^2$.
Then, we derive the parameters of the scalar potential
from eqs.~\eqref{reconstruct_lambdas},
and maintain those points which provide a bounded from below
solution \cite{Deshpande:1977rw},
conforming to perturbative unitarity
\cite{Kanemura:1993hm, Akeroyd:2000wc,Ginzburg:2003fe},
and the oblique radiative parameters $S, T, U$
\cite{Grimus:2008nb, Baak:2012kk}.
After implementing this algorithm,
we have a collection of possible C2HDM data points.

We generate the rates for all channels,
including all production mechanisms.
We use the expressions in the appendices,
and utilize HIGLU \cite{Spira:1995mt} at NNLO for
$gg \ra h$  (gluon fusion),
SusHi \cite{Harlander:2012pb} at NNLO for
$b \bar{b} \ra h$,
and ref.~\cite{LHCCrossSections} for
$Vh$ (associated production),
$t \bar{t} h$,
and $VV \ra h$ (vector boson fusion).
The expressions for the decay rates are obtained
in the appendices.
In particular,
$h \ra Z \gamma$ is explained in great detail
in \ref{app:hZgamma} and \ref{app:hZgamma_widths},
for a generic scalar/pseudoscalar mixed state $h$.
Finally,
we compute the ratio of rates
\be
\mu_f
=
\frac{\sigma^\textrm{2HDM}(pp \ra h)}{\sigma^\textrm{SM}(pp \ra h)}\,
\frac{\Gamma^\textrm{2HDM}[h \ra f]}{\Gamma^\textrm{SM}[h \ra f]}\,
\frac{\Gamma^\textrm{SM}[h \ra \textrm{all}]}{
\Gamma^\textrm{2HDM}[h \ra \textrm{all}]}\, ,
\label{mus}
\ee
where $\sigma$ is the cross section for Higgs production,
$\Gamma[h \ra f]$ is the decay width into the final state $f$,
and $\Gamma[h \ra \textrm{all}]$ is $h$'s total width.
The ratios $\mu_f$ can then be compared with those
quoted by the experimental collaborations.
For definiteness,
our discussions will be based on the ATLAS
\cite{atlas:ichep2014} and CMS \cite{cms:ichep2014}
results presented in the plenary talks at ICHEP2014,
which we summarize in table~\ref{tab:2}.
%
%
\begin{table}[tbp]
\centering
\begin{tabular}{|ccccc|}
\hline
channel & & ATLAS  & & CMS  \\
\hline
$\mu_{\gamma\gamma}$  & &
$1.57^{+0.33}_{-0.28}$   & &
$1.13 \pm 0.24$
\\*[2mm]
$\mu_{WW}$  & &
$1.00^{+0.32}_{-0.29}$   & &
$0.83 \pm 0.21$
\\*[2mm]
$\mu_{ZZ}$  & &
$1.44^{+0.40}_{-0.35}$   & &
$1.00 \pm 0.29$
\\*[2mm]
$\mu_{\tau^+\tau^-}$  & &
$1.4^{+0.5}_{-0.4}$   & &
$0.91 \pm 0.27$
\\*[2mm]
$\mu_{b \bar{b}}$  & &
$0.2^{+0.7}_{-0.6}$   & &
$0.93 \pm 0.49$
\\*[2mm]
\hline
\end{tabular}
\caption{\label{tab:2} Experimental results presented by
ATLAS and CMS at ICHEP2014.}
\end{table}
Notice that the errors are still important;
combining ATLAS and CMS would lead to errors
of order $20\%$ in $VV$, $\gamma\gamma$,
and slightly larger in $\tau^+ \tau^-$.
On the other hand,
the errors on $b \bar{b}$,
which is only detected in associated production,
are of order $50\%$.
In particular,
ATLAS excludes the SM $\mu_{\gamma\gamma} = 1$
($\mu_{ZZ} = 1$)
at 2-$\sigma$ (1-$\sigma$),
while CMS is within 1-$\sigma$ of the SM on all channels.

\subsection{Type I model}

To study the effect of current experimental bounds on the pseudoscalar
content of the 125 GeV Higgs,
we follow ref.~\cite{pseudo} and study three sets of points:
points where the $h_1$ is mainly scalar,
with $|s_2| < 0.1$ (in green/light-grey in the simulation
figures to be shown below);
points where the $h_1$ is mainly pseudoscalar,
with $|s_2| > 0.85$ (in red/dark-grey in the simulation
figures to be shown below);
points where the $h_1$ is a almost even mix
of scalar and pseudoscalar,
with $0.45 < |s_2| < 0.55$ (in blue/black in the simulation
figures to be shown below).

To compare with current experiments,
all figures in this article will be drawn for processes at 8 TeV,
except were noted otherwise.
The exceptions are figures drawn at 14 TeV,
designed to foresee future experimental reaches.
Nevertheless,
we have checked that there are very small differences
between 8 TeV and 14 TeV,
for the figures that interest us.
As explained in \cite{Fontes:2014tga},
this is due to the fact that the ratio between the dominant
and sub-dominant gluon fusion production mechanisms
(which, in the two Higgs doublet model, can be relevant
with both top and bottom quarks in the loop)
remains very similar as one changes from
8 TeV to 14 TeV in our HIGLU simulations.

\begin{figure}[tbp]
\centering 
\includegraphics[width=.45\textwidth]{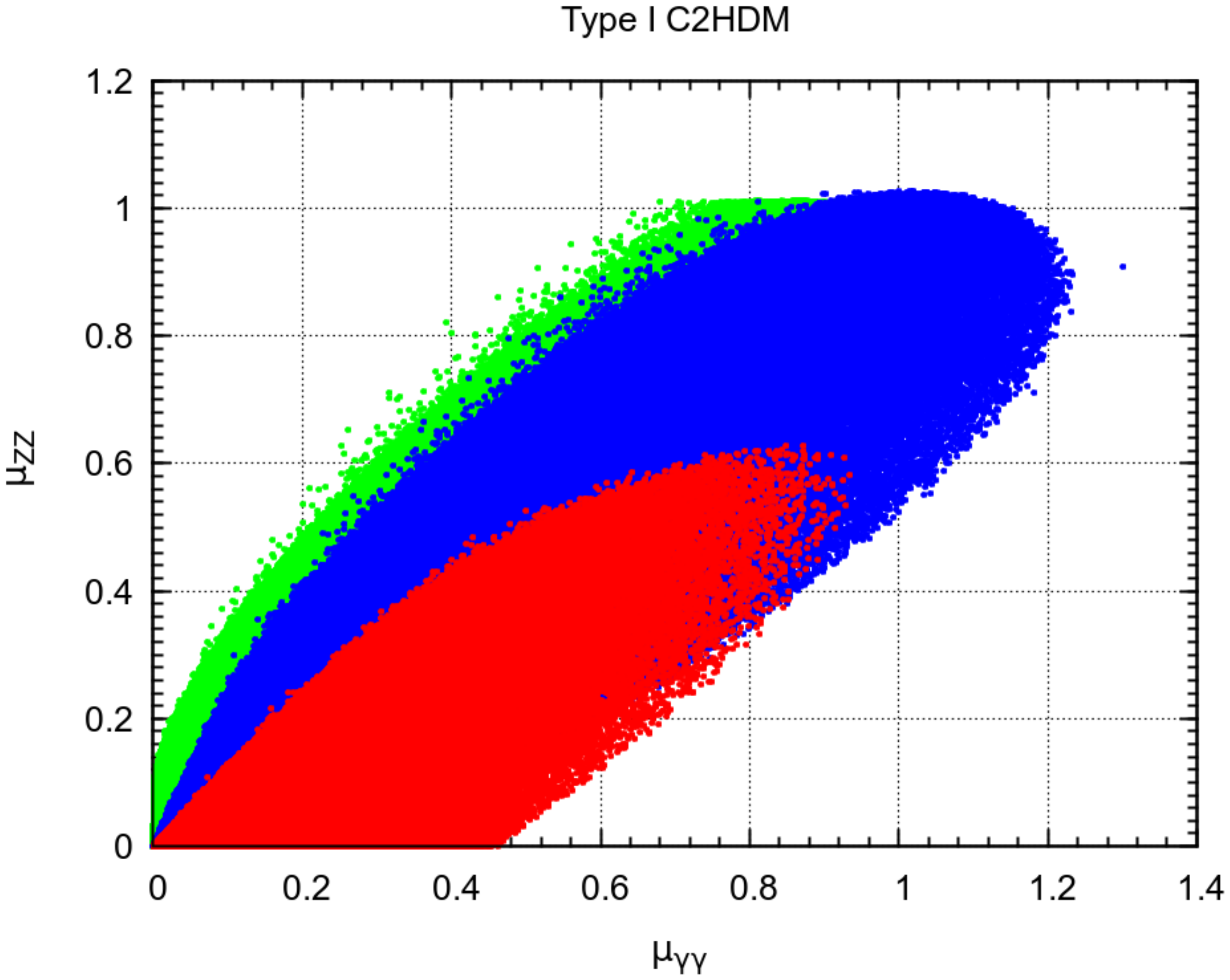}
\hfill
\includegraphics[width=.45\textwidth]{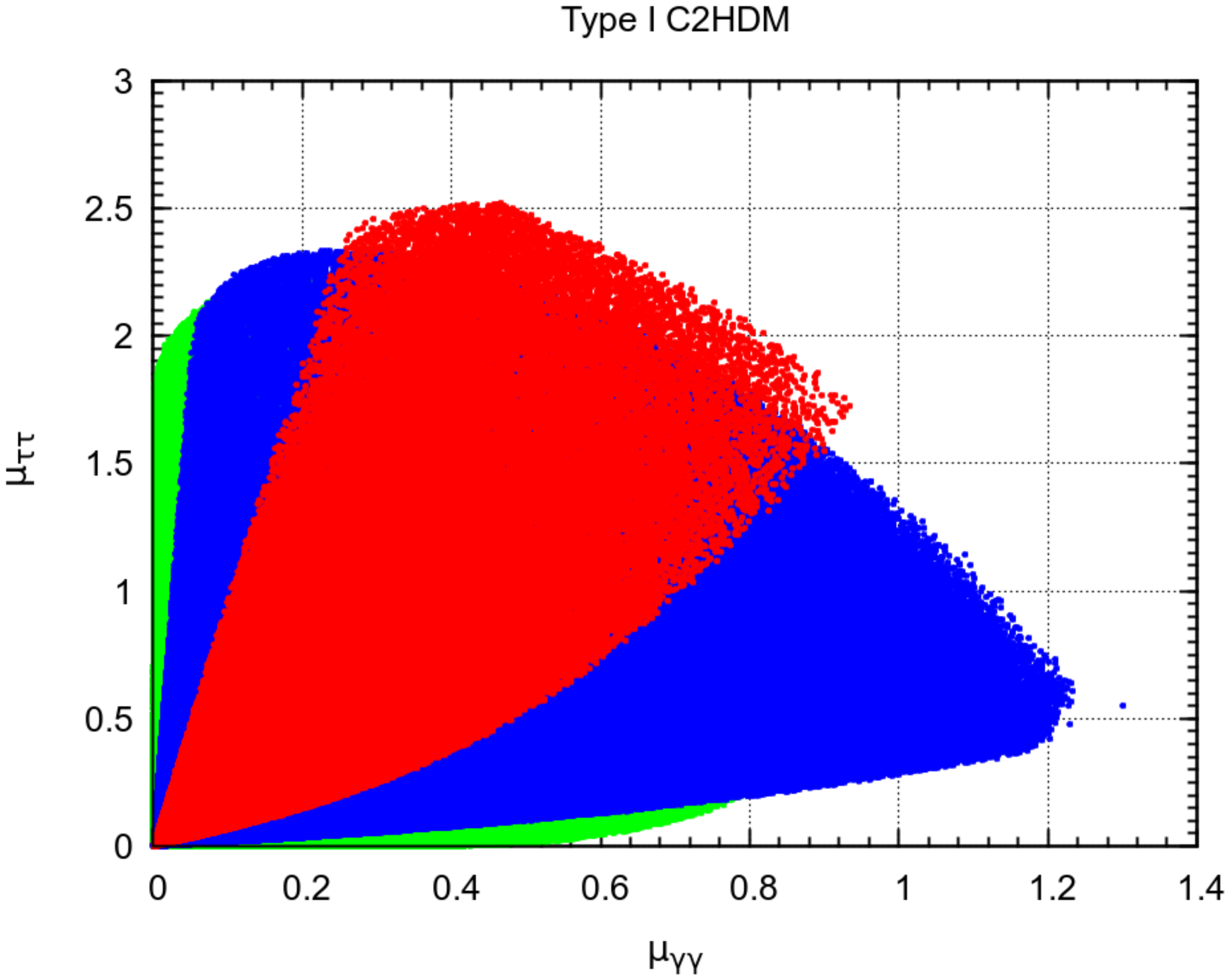}
\caption{\label{fig:typeI_ZZ_tautau} Left panel:
Results in the $\mu_{ZZ}$ - $\mu_{\gamma\gamma}$ plane
(left panel)
and in the $\mu_{\tau^+\tau^-}$ - $\mu_{\gamma\gamma}$ plane
(right panel)
for the Type I C2HDM.
The points in green/light-grey,
blue/black,
and red/dark-grey
correspond to $|s_2| < 0.1$,
$0.45 < |s_2| < 0.55$,
and $|s_2| > 0.85$,
respectively.}
\end{figure}

Our results for $\mu_{ZZ}$ versus $\mu_{\gamma\gamma}$
are shown in the left panel of fig.~\ref{fig:typeI_ZZ_tautau}.
This can be compared with fig.~1 of ref.~\cite{pseudo}.
We get qualitatively the same results,
meaning that $|s_2| > 0.85$ is excluded by CMS at 1-$\sigma$.
Also, larger values of $\mu_{\gamma\gamma}$
are obtained with $0.45 < |s_2| < 0.55$ than with $|s_2| < 0.1$.
Thus, a putative future result of,
for example,
$\mu_{\gamma\gamma} = 1.3 \pm 0.1$
(consistent with the current ATLAS bound)
would imply that
the Higgs found at LHC has comparable scalar and pseudoscalar
components.
Notice from the left panel of fig.~\ref{fig:typeI_ZZ_tautau}
that this would be consistent with $\mu_{ZZ} \sim 0.9$
but less so with $\mu_{ZZ} \sim 1$.

On the right panel of fig.~\ref{fig:typeI_ZZ_tautau},
we show our results
in the $\mu_{\gamma\gamma} - \mu_{\tau^+\tau^-}$ plane.
This can be compared with fig.~2 of ref.~\cite{pseudo}
which shows $\mu_{b \bar{b}}$ considering,
as we correct below, all production channels.
There is qualitative agreement, but there are subtle differences,
because we are using the latest version of HIGLU \cite{Spira:1995mt},
and, eventually, different PDF's and energy scales.
The difference is apparent when plotting $\mu_{\tau^+\tau^-}$
as a function of $\tan{\beta}$.
As shown in ref.~\cite{Fontes:2014tga},
$\mu_{\tau^+\tau^-}$ is
very sensitive to the production rates
(and, thus, should be interpreted with care),
while $\mu_{\gamma\gamma}$  and $\mu_{Z\gamma}$ are not.
With this caution,
we find that values as large as $\mu_{\tau^+\tau^-} \sim 2$
are allowed.
If one requires $\mu_{\gamma\gamma} \sim 1$,
then $\mu_{\tau^+\tau^-}$ lies roughly between
$0.4$ and $1.4$.

\begin{figure}[tbp]
\centering 
\includegraphics[width=.45\textwidth]{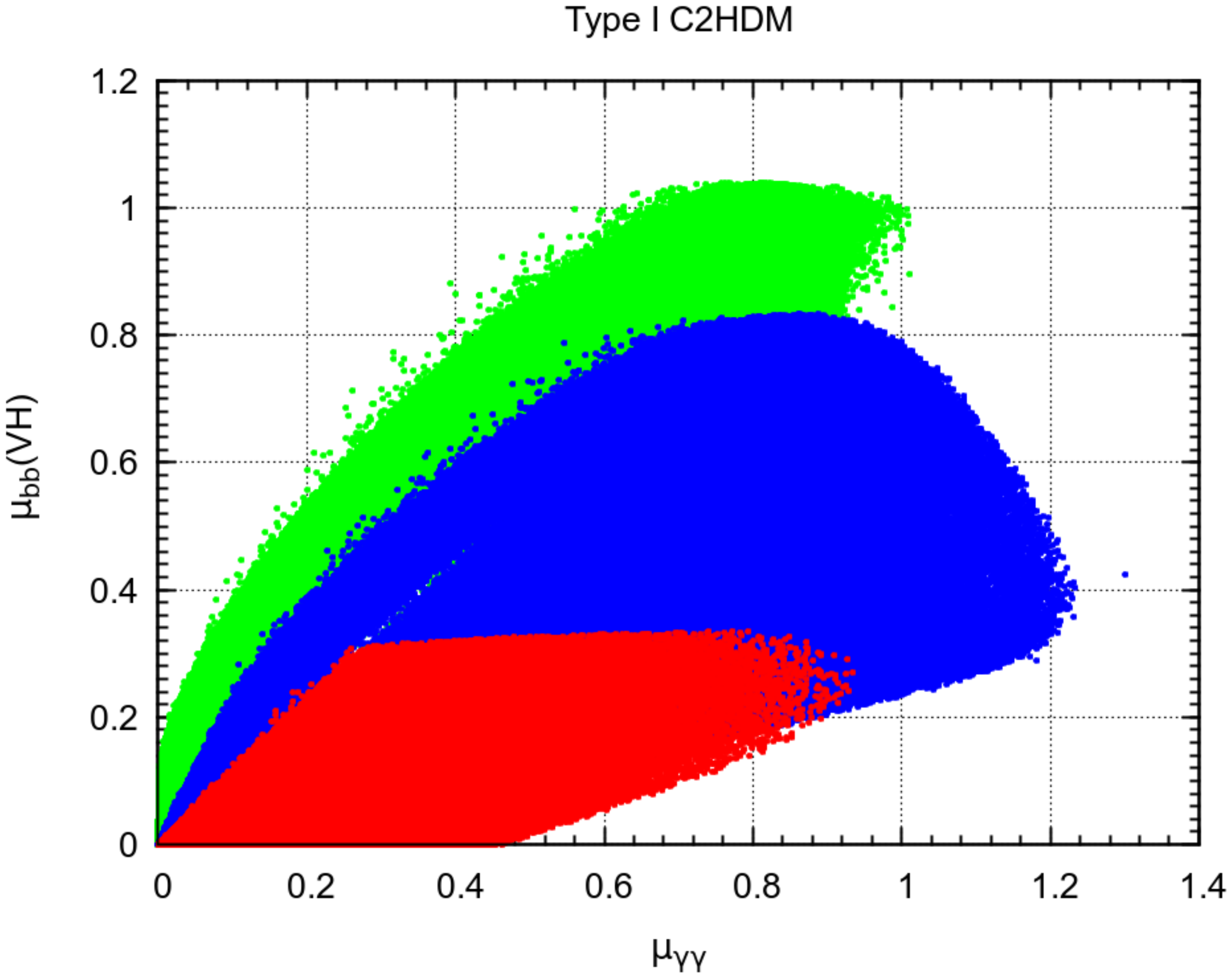}
\hfill
\includegraphics[width=.45\textwidth]{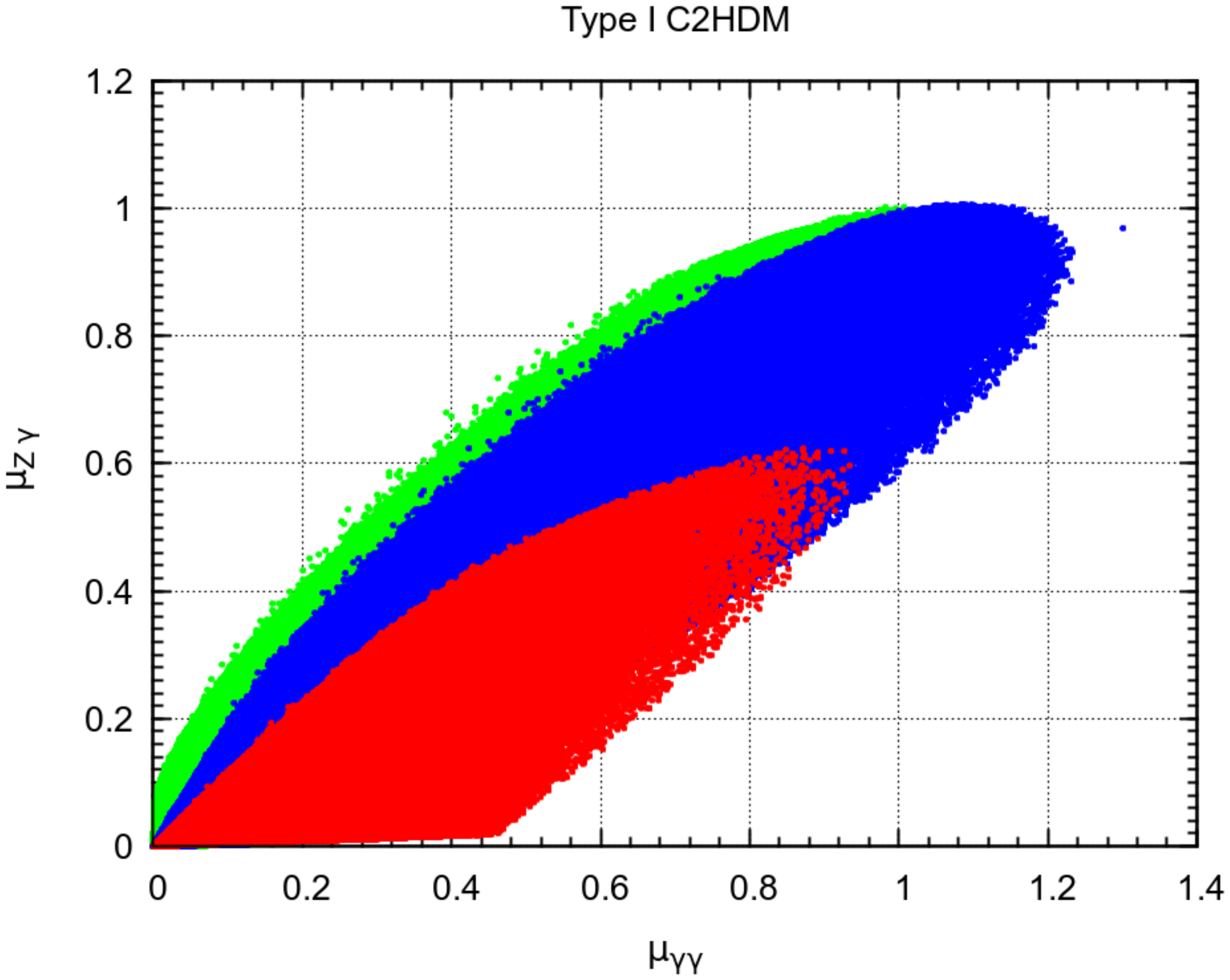}
\caption{\label{fig:typeI_bb_Zph} Left panel:
Results in the $\mu_{b\bar{b}}(Vh)$ - $\mu_{\gamma\gamma}$ plane
(left panel)
and in the $\mu_{Z \gamma}$ - $\mu_{\gamma\gamma}$ plane
(right panel)
for the Type I C2HDM.
The points in green/light-grey,
blue/black,
and red/dark-grey
correspond to $|s_2| < 0.1$,
$0.45 < |s_2| < 0.55$,
and $|s_2| > 0.85$,
respectively.}
\end{figure}

In ref.~\cite{pseudo},
$\mu_{b \bar{b}}$ was calculated using all
production channels.
Here we use exclusively the $Vh$ production mechanism that
allows detection at LHC.
Our results are shown on the left panel of fig.~\ref{fig:typeI_bb_Zph}.
In the Type I model,
$\mu_{b\bar{b}}(Vh) \lesssim 1.1$ for all values of $|s_2|$,
and $\mu_{b\bar{b}}(Vh) \lesssim 0.35$ for $|s_2| > 0.85$.
Thus,
we learn that CMS excludes again $|s_2| > 0.85$ at 1-$\sigma$
(recall that even the SM $ZZ$ and $\gamma\gamma$ are outside
ATLAS' 1-$\sigma$ intervals),
and a good measurement of $\mu_{b\bar{b}}(Vh)$ will be useful
in ruling out large pseudoscalar components.

\begin{figure}[tbp]
\centering 
\includegraphics[width=.45\textwidth]{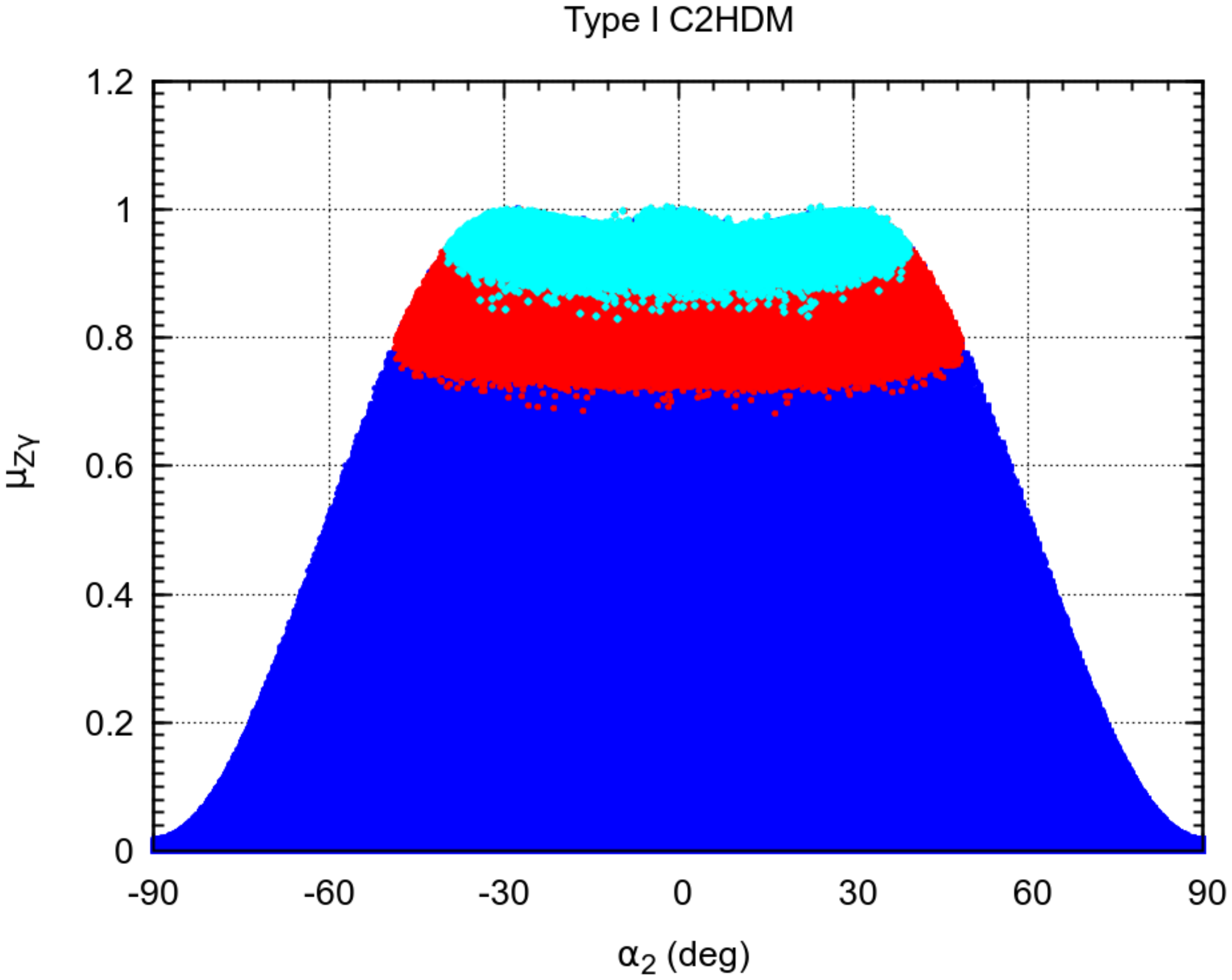}
\hfill
\includegraphics[width=.45\textwidth]{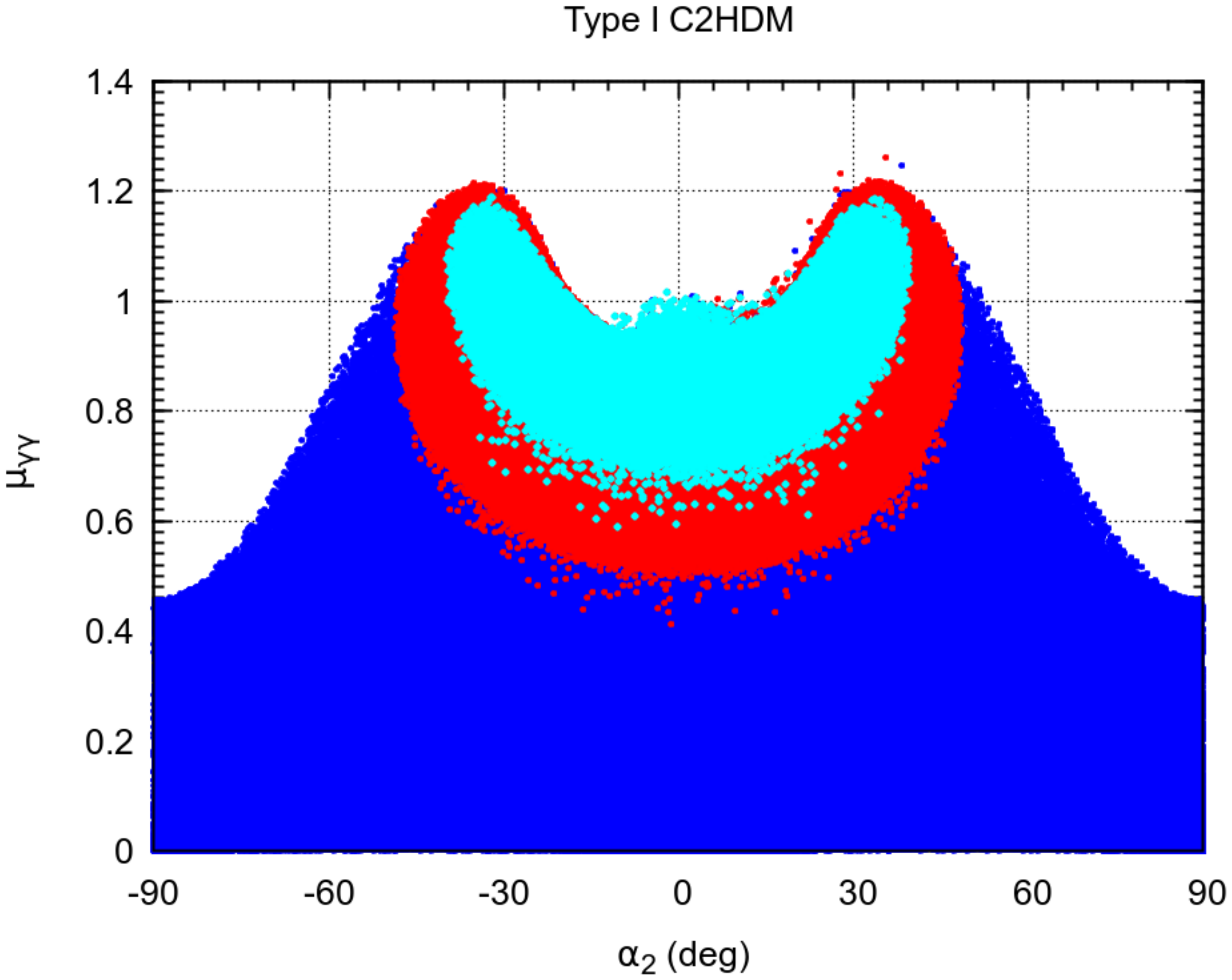}
\caption{\label{fig:typeI_Zph_phph_s2} Figures of
$\mu_{Z\gamma}$ ($\mu_{\gamma\gamma}$) on the
left (right) panel,
as a function of $s_2$.
The points in red/dark-grey (cyan/light-grey)
where chosen to obey $\mu_{VV} = 1$ within
20\% (5\%).
These figures have been drawn for 14 TeV.}
\end{figure}

Now we turn to one of the main motivations for this work.
The right panel of fig.~\ref{fig:typeI_bb_Zph} shows our results
in the $\mu_{\gamma\gamma} - \mu_{Z\gamma}$ plane.
We notice that large pseudoscalar components
(large $|s_2|$) imply small values for $\mu_{Z\gamma}$.
There are two points to stress.
First,
there is a strong correlation between $\mu_{Z \gamma}$
and $\mu_{\gamma \gamma}$, even when all values of
$s_2$ are taken into account.
Second,
that correlation is partly connected with $s_2$.
This can be seen in the blue/black regions of
figs.~\ref{fig:typeI_Zph_phph_s2},
where we see that large values of
$\mu_{Z\gamma}$ and $\mu_{\gamma\gamma}$ are only
possible around $s_2 \sim 0$ and $h_1$ with a large
scalar component.
In contrast,
a large pseudoscalar component implies very small values
for both $\mu_{Z\gamma}$ and $\mu_{\gamma\gamma}$.
As a result,
a value of $\mu_{Z \gamma} \sim 1$ would be very efficient in ruling
out a large pseudoscalar component.
Figs.~\ref{fig:typeI_Zph_phph_s2}
also show in red/dark-grey (cyan/light-grey) the allowed regions
if we assume that the measurements of $\mu_{VV}$ at 14 TeV
will center around unity with a $20\%$ ($5\%$) error.
The $VV$ constraint implies that $\mu_{\gamma\gamma}$ and
$\mu_{Z\gamma}$ are expected to lie close to their SM value
in the C2HDM and that $|\alpha_2|$ should lie below
$50$ degrees.
A similar analysis of the impact of $VV$,
shows that $\alpha_3$ can take any value and that
$|\alpha_1|$ should be larger than about $60$ degrees.

\subsection{Type II model}

\begin{figure}[tbp]
\centering 
\includegraphics[width=.45\textwidth]{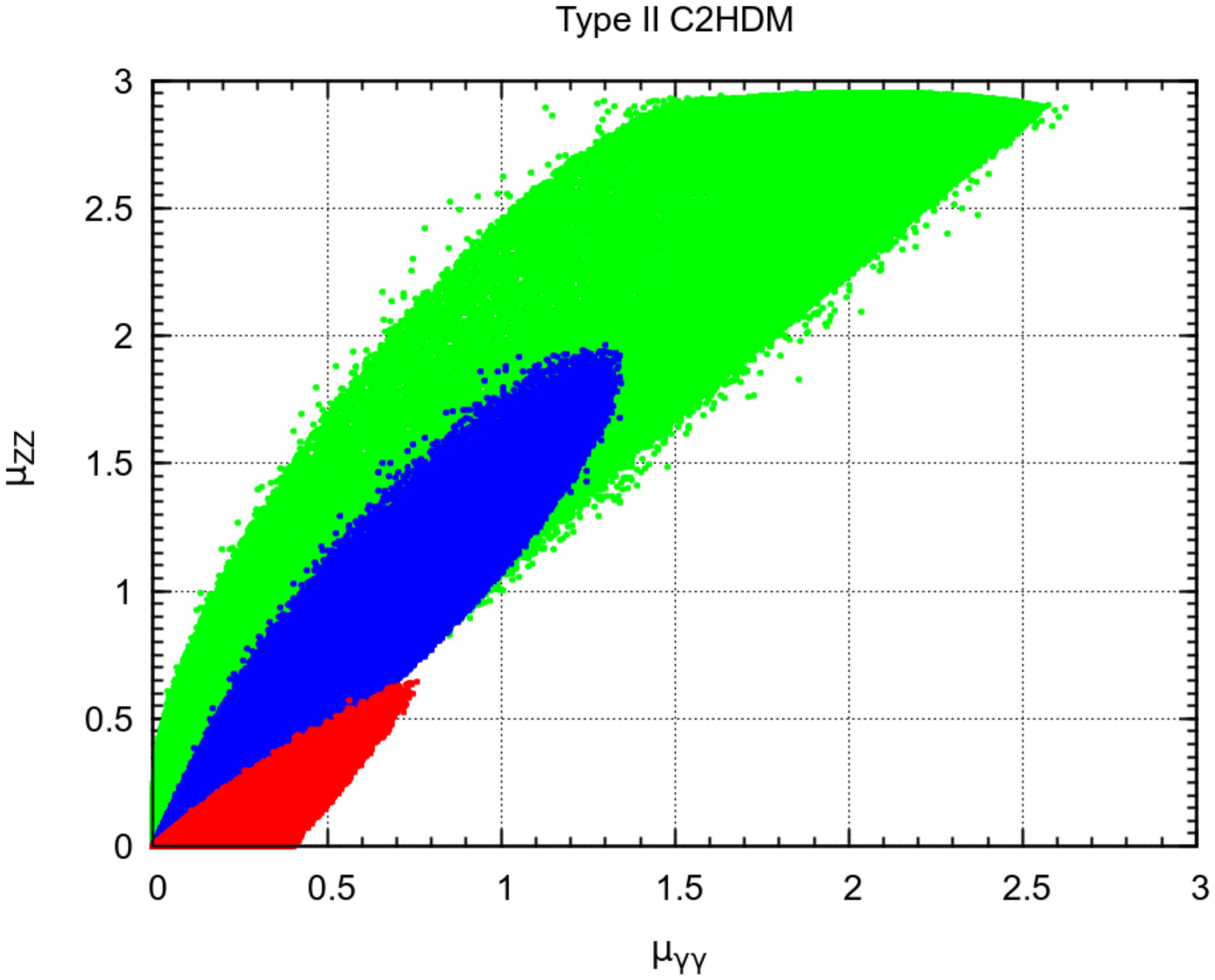}
\hfill
\includegraphics[width=.45\textwidth]{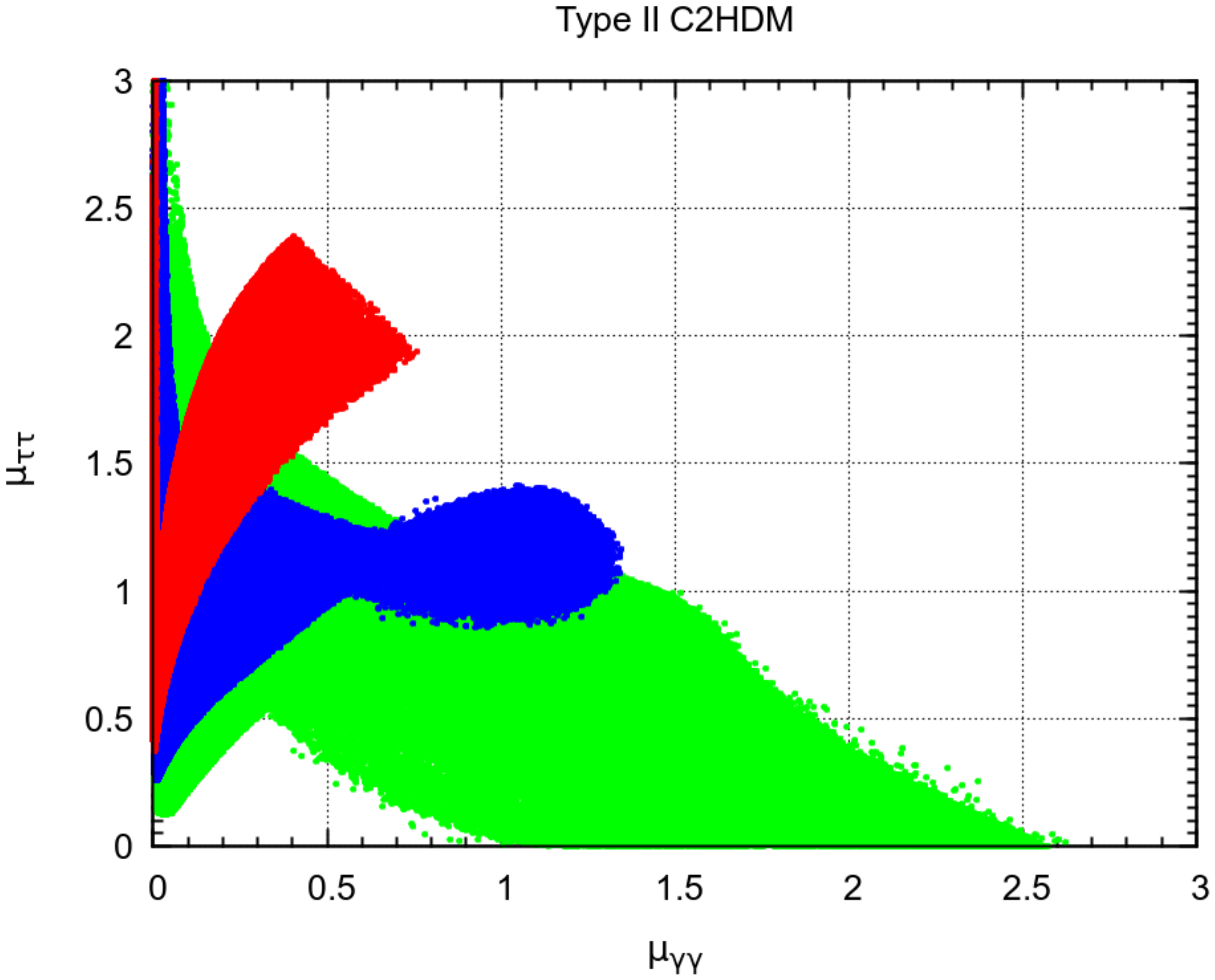}
\caption{\label{fig:typeII_ZZ_tautau} Left panel:
Results in the $\mu_{ZZ}$ - $\mu_{\gamma\gamma}$ plane
(left panel)
and in the $\mu_{\tau^+\tau^-}$ - $\mu_{\gamma\gamma}$ plane
(right panel)
for the Type II C2HDM.
The points in green/light-grey,
blue/black,
and red/dark-grey
correspond to $|s_2| < 0.1$,
$0.45 < |s_2| < 0.55$,
and $|s_2| > 0.85$,
respectively.}
\end{figure}

The results obtained in Type II for $\mu_{ZZ}$ versus $\mu_{\gamma\gamma}$
are shown in the left panel of fig.~\ref{fig:typeII_ZZ_tautau}.
In this model, values as large as
$\mu_{\gamma\gamma} \sim 2.5$
and
$\mu_{ZZ} \sim 3$ are allowed for small values of $s_2$.
In contrast,
$|s_2| > 0.85$ forces both to be smaller than $0.8$.
This means that even the high central values quoted by ATLAS
are consistent with a Type II C2HDM where $h_1$ has a
dominant scalar component.
In fact,
one can find $s_2 < 0.1$ but also a few
$0.45 < |s_2| < 0.55$ points within
the ATLAS and CMS 1-$\sigma$ bounds.
As occurred in Type I,
both experiments exclude a large
pseudoscalar component
($|s_2| > 0.85$) at more than 1-$\sigma$.
However,
in contrast to Type I,
here the largest values of $\mu_{\gamma\gamma} $
occur for $s_2 < 0.1$ and not for
$0.45 < |s_2| < 0.55$.
That is,
in Type I a large value ($\mu_{\gamma\gamma} \sim 1.2$)
favors a comparable scalar/pseudoscalar mix,
while in Type II a large value (here, $\mu_{\gamma\gamma} \geq 1.2$)
favors a pure scalar.

Curiously,
the situation is the reverse when one considers
$\mu_{\tau^+ \tau^-}$,
which we show on the right panel of fig.~\ref{fig:typeII_ZZ_tautau}.
For example,
for $\mu_{\gamma\gamma} \sim 1$,
a value of $\mu_{\tau^+ \tau^-} \sim 1.3$
favors an even scalar/pseudoscalar mix
over the pure scalar solution.
In contrast,
$|s_2|$ is less easily constrained from
$\mu_{b\bar{b}}(Vh)$,
although $\mu_{b\bar{b}}(Vh) \gtrsim 0.4$
rules out $|s_2|>0.85$.
Looking at the various channels,
both CMS and ATLAS rule out
$|s_2|>0.85$ by more than 2-$\sigma$ in Type II C2HDM.
Better measurements of $\gamma\gamma$,
$\tau^+\tau^-$,
and $b\bar{b} (Vh)$ will be instrumental in
determining $s_2$.

\begin{figure}[tbp]
\centering 
\includegraphics[width=.45\textwidth]{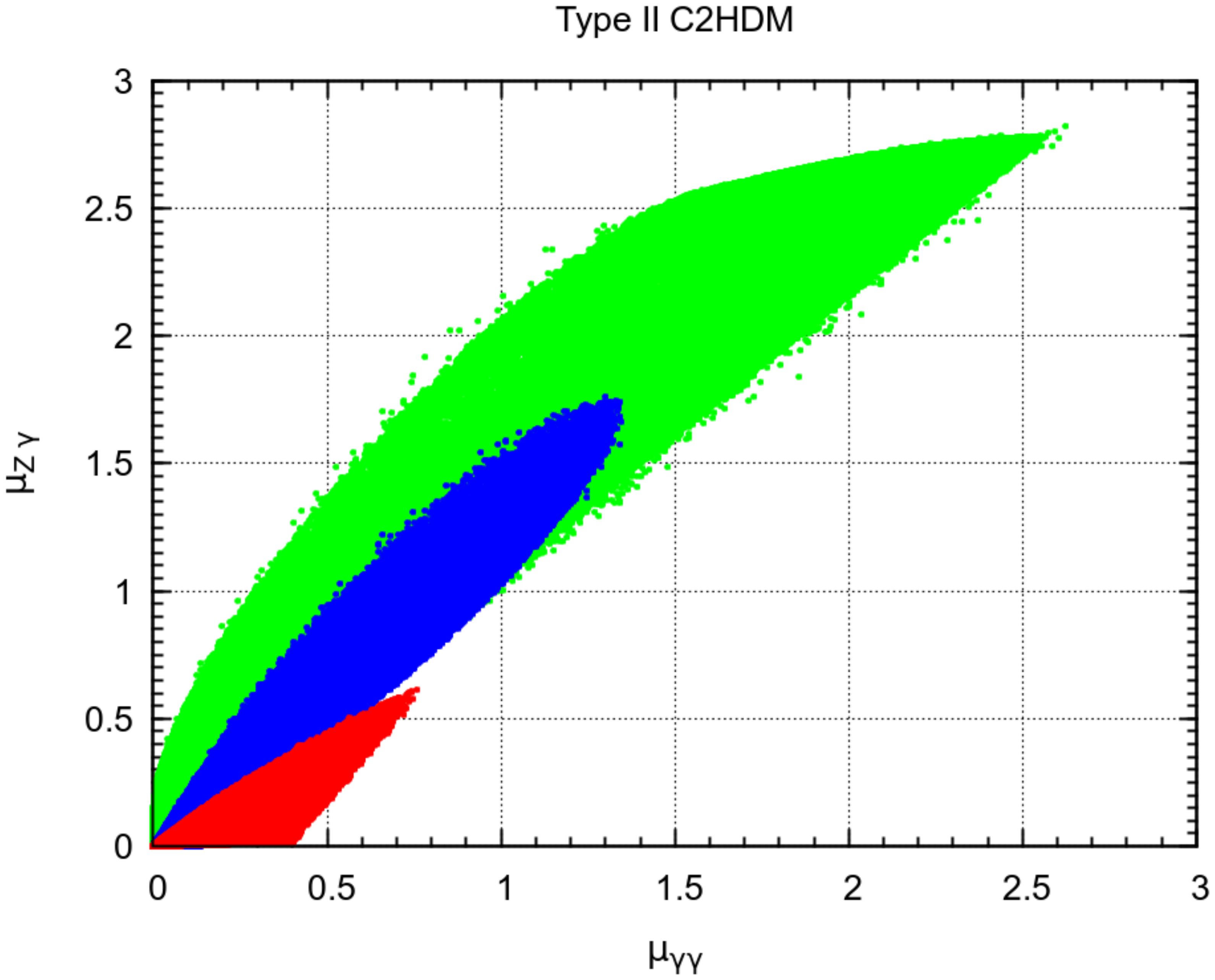}
\hfill
\includegraphics[width=.45\textwidth]{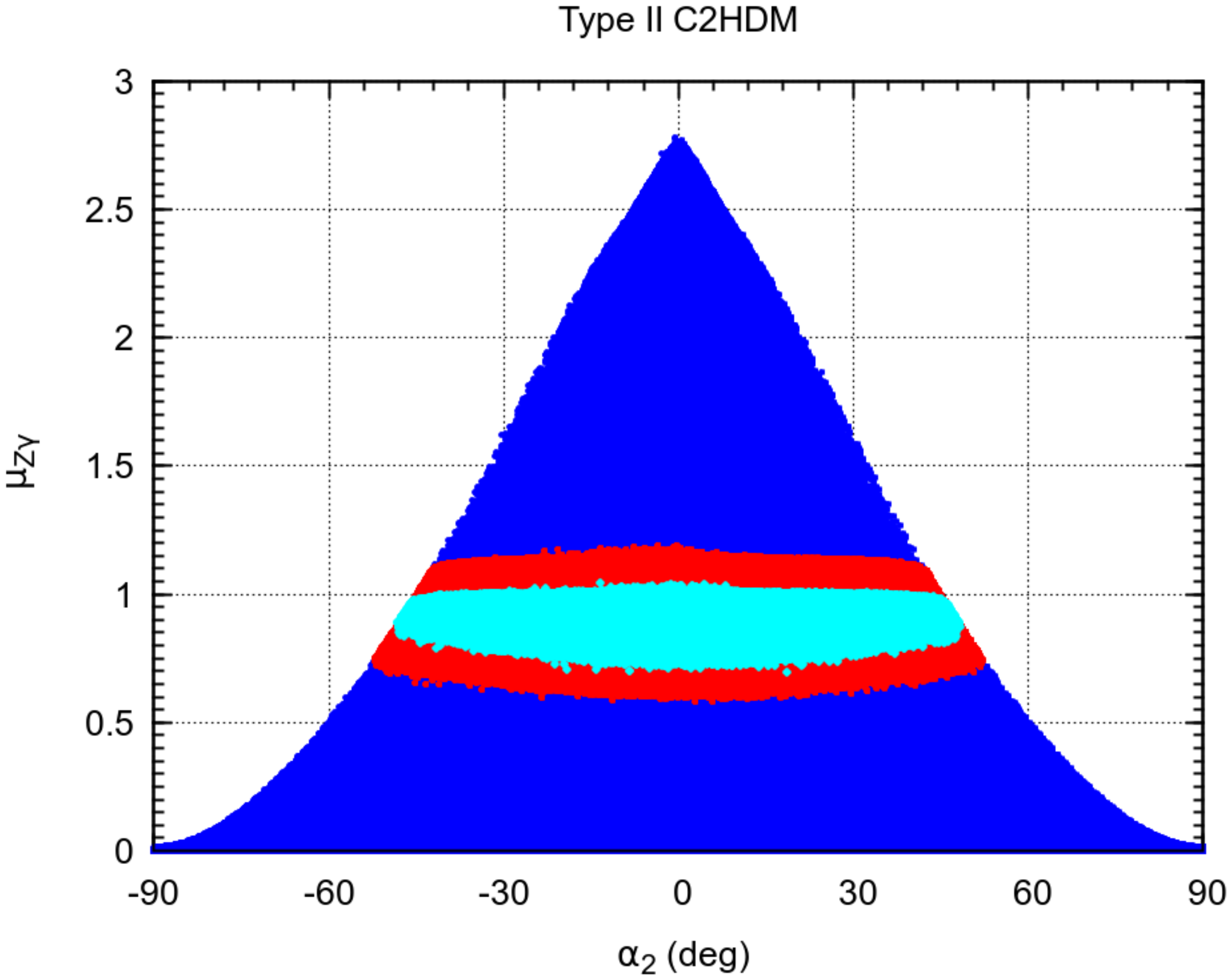}
\caption{\label{fig:typeII_Zph} Left panel:
Type II results in the $\mu_{Z\gamma}$ - $\mu_{\gamma\gamma}$ plane.
The points in green/light-grey,
blue/black,
and red/dark-grey
correspond to $|s_2| < 0.1$,
$0.45 < |s_2| < 0.55$,
and $|s_2| > 0.85$,
respectively.
Right panel:
Type II predictions in the $\mu_{Z\gamma}$ - $s_2$ plane.
The points in red/dark-grey (cyan/light-grey)
where chosen to obey $\mu_{VV} = 1$ within
20\% (5\%). This figure has been draw at 14 TeV.
}
\end{figure}

Next,
we consider the simulations for $Z\gamma$,
shown in on the left panel of
fig.~\ref{fig:typeII_Zph}.
Large values for $\mu_{Z \gamma}$
are possible for small $|s_2|$.
Comparing with the right panel
of fig.~\ref{fig:typeI_bb_Zph} we see that
in Type II much larger values of $\mu_{Z \gamma}$
(and of $\mu_{\gamma \gamma}$) are allowed,
but that there is still a strong correlation
between the two which,
again,
is partly due to $s_2$.
This is shown on the right panel of fig.~\ref{fig:typeII_Zph},
where we see that large values of $\mu_{Z\gamma}$ require
large values of $\mu_{VV}$ and correspond to an almost
pure scalar.
Measurements of $\mu_{VV}$ within $20\%$ of unity,
force $\mu_{Z\gamma} \sim 1$ and require $|\alpha_2| \lesssim 50$
degrees.
This puts a further bound on a large pseudoscalar component.

\subsection{Lepton Specific model}

In this case,
the results for $\mu_{ZZ}$ and $\mu_{b\bar{b}}(Vh)$
versus $\mu_{\gamma\gamma}$ are very similar to those
presented on the left panels of figs.~\ref{fig:typeI_ZZ_tautau}
and \ref{fig:typeI_bb_Zph} for Type I,
respectively.
The same holds for $\mu_{Z\gamma}$,
shown on the right panel of fig.~\ref{fig:typeI_bb_Zph}.
Minute differences are as follows.
Close to $\mu_{\gamma\gamma} \sim 1$,
one can get slightly larger values for $\mu_{ZZ}$,
up to approximately $1.1$.
Conversely,
$\mu_{\gamma\gamma} \lesssim 1.1$ here,
while $\mu_{\gamma\gamma} \lesssim 1.3$
in Type I.
Here,
as in Type I,
$|s_2| > 0.85$ forces $\mu_{b\bar{b}}(Vh) < 0.3$.
Thus, a good measurement of $\mu_{b\bar{b}}(Vh)$
will be instrumental in ruling out large pseudoscalar components.

\begin{figure}[tbp]
\centering 
\includegraphics[width=.45\textwidth]{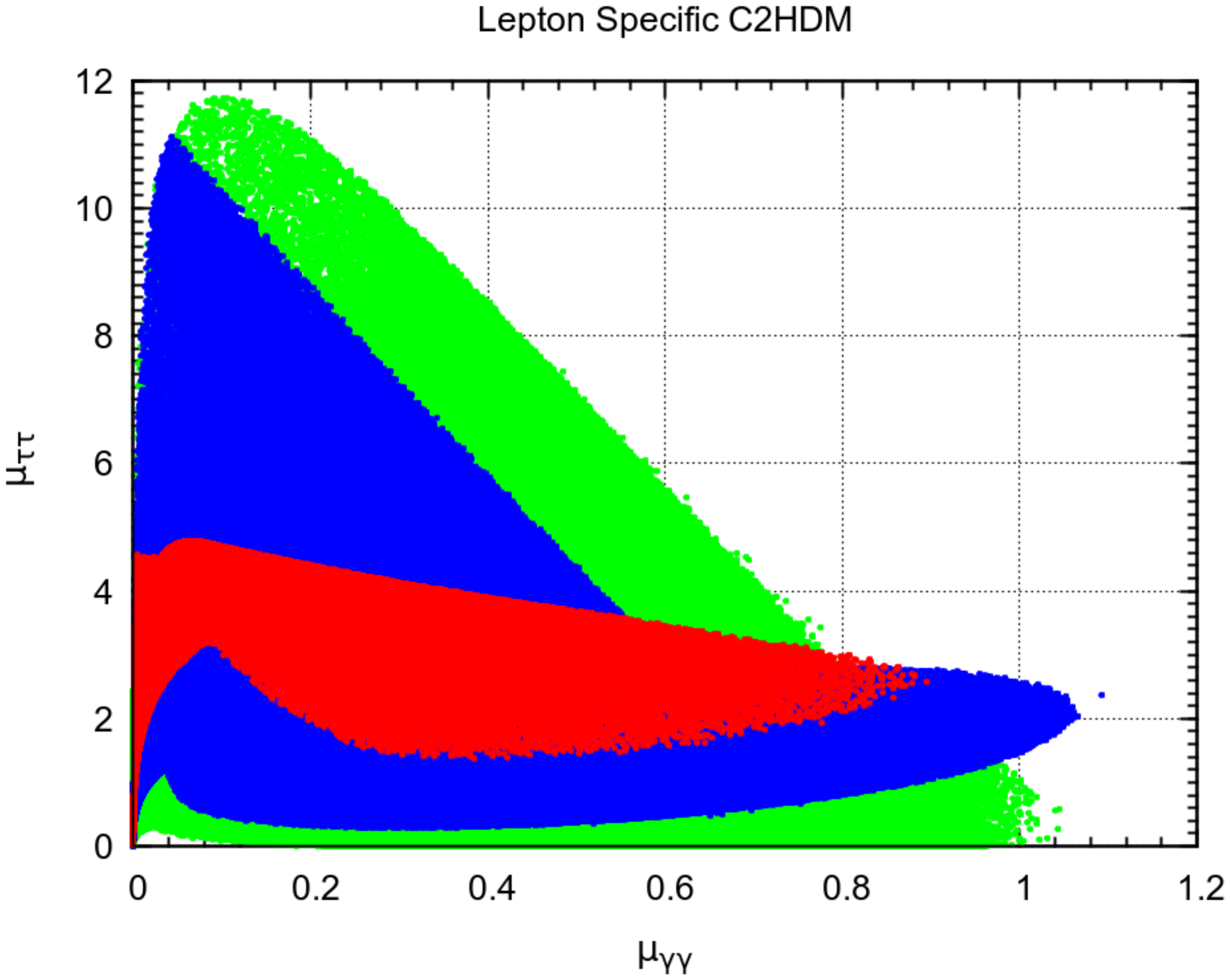}
\caption{\label{fig:LS_tautau} Lepton Specific
simulations in the $\mu_{\tau^+\tau^-}$ - $\mu_{\gamma\gamma}$ plane.
The points in green/light-grey,
blue/black,
and red/dark-grey
correspond to $|s_2| < 0.1$,
$0.45 < |s_2| < 0.55$,
and $|s_2| > 0.85$,
respectively.
}
\end{figure}

As expected,
the situation for $\mu_{\tau^+\tau^-}$ differs,
as shown in fig.~\ref{fig:LS_tautau}.
A large pseudoscalar component
($|s_2|>0.85$)
forces $\mu_{\tau^+\tau^-} > 1.2 $ when $\mu_{\gamma\gamma} > 0.1$.
These values are ruled out by CMS at 1-$\sigma$.
ATLAS, on the other hand,
is barely consistent with these values for
$\mu_{\tau^+\tau^-}$,
but rules out this model (and the SM)
in $\mu_{\gamma\gamma}$ at 1-$\sigma$.

\subsection{Flipped model}

The results for $\mu_{\gamma\gamma}$,
$\mu_{ZZ}$,
$\mu_{b\bar{b}}(Vh)$,
and $\mu_{Z\gamma}$
in this model, are similar
to those for Type II.
Slight differences are as follows.
Here $\mu_{\gamma\gamma}$ ($\mu_{ZZ}$, $\mu_{\gamma\gamma}$ )
can only be as large as $2.2$
($2.5$, $2.4$),
while one could achieve $2.5$ ($2.9$, $2.8$) in Type II.
The situation for $\mu_{b\bar{b}}(Vh)$ is virtually the same.
In particular,
$|s_2|>0.85$ is ruled out at 1-$\sigma$ by both
ATLAS and CMS.

\begin{figure}[tbp]
\centering 
\includegraphics[width=.45\textwidth]{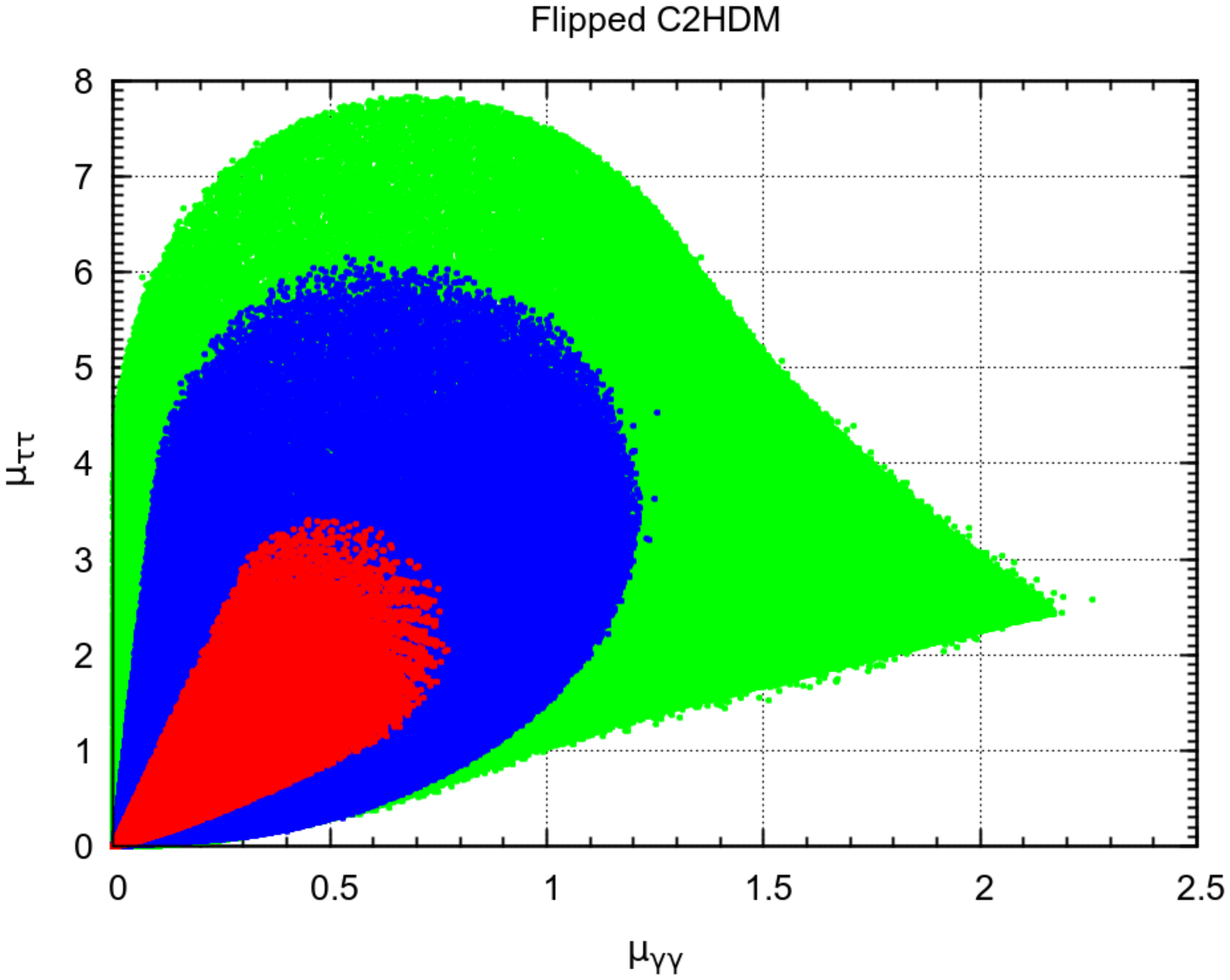}
\hfill
\includegraphics[width=.45\textwidth]{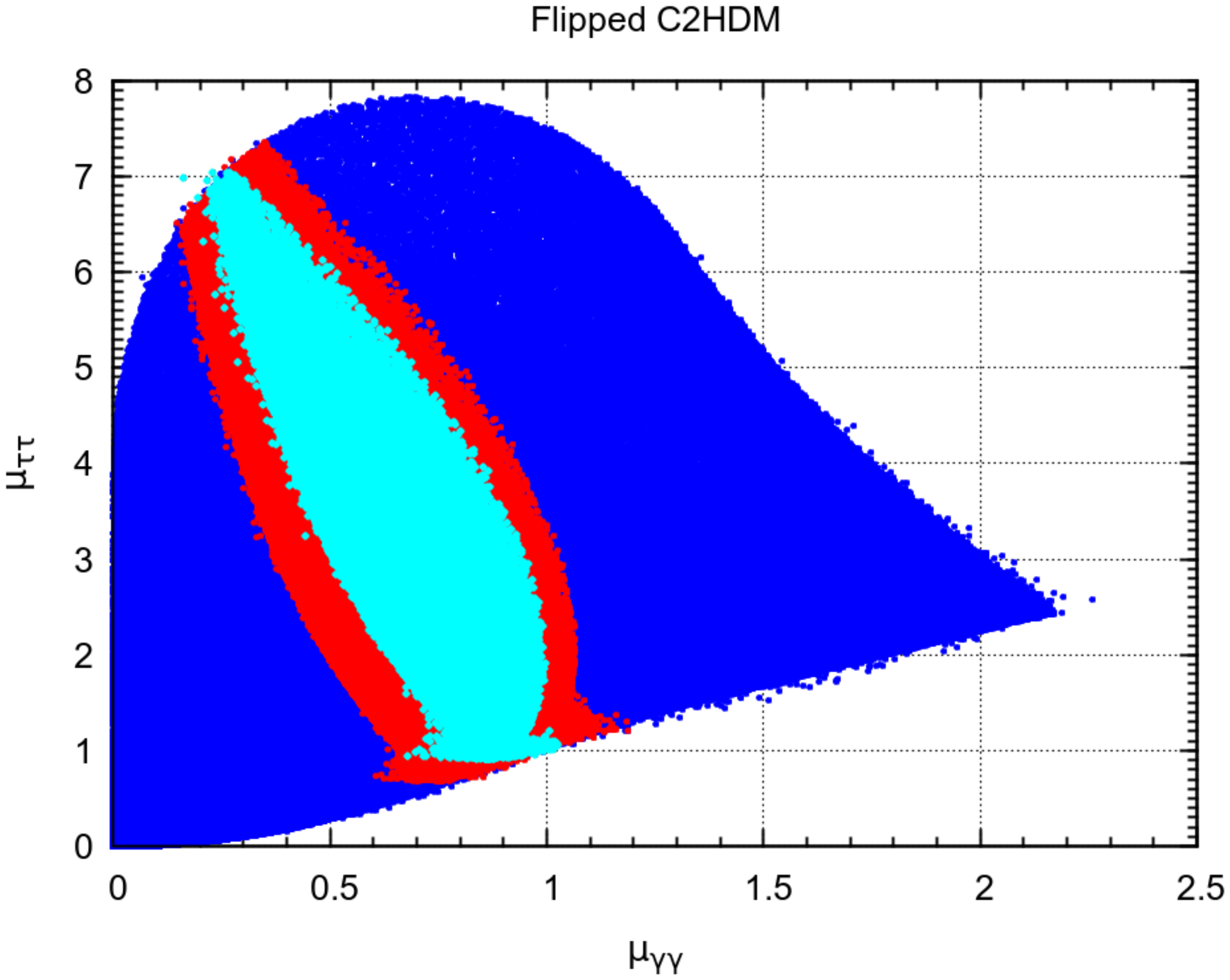}
\caption{\label{fig:Fli_tautau} Left panel:
Flipped model
results in the $\mu_{\tau^+\tau^-}$ - $\mu_{\gamma\gamma}$ plane.
The points in green/light-grey,
blue/black,
and red/dark-grey
correspond to $|s_2| < 0.1$,
$0.45 < |s_2| < 0.55$,
and $|s_2| > 0.85$,
respectively.
Right panel:
same as left,
except that all values for
$s_2$ are included as blue/black points.
Also shown as red/dark-grey (cyan/light-grey)
are those points which obey $\mu_{VV} = 1$ within
20\% (5\%).
}
\end{figure}

The situation is very different for
$\mu_{\tau^+\tau^-}$,
as shown on the left panel of fig.~\ref{fig:Fli_tautau}.
Notice that one can find points as large as
$\mu_{\tau^+\tau^-} = 7.5$ for reasonable values of
$\mu_{\gamma\gamma} \sim 1$.

As mentioned in ref.~\cite{Fontes:2014tga},
constraints on $\mu_{VV}$ have a very strong
impact on predictions in Type II and Flipped models,
which have a simple trigonometric interpretation.
One might wonder whether large values for
$\mu_{\tau^+\tau^-}$ are consistent
with $\mu_{VV}$.
This is shown on the right panel of fig.~\ref{fig:Fli_tautau}:
the red/dark-grey (cyan/light-grey)
are those points which obey $\mu_{VV} = 1$ within
20\% (5\%).
We see that large values of $\mu_{\tau^+\tau^-}$
are still allowed.
Thus,
$\mu_{\tau^+\tau^-}$ will have an enormous impact in probing
the Flipped C2HDM.

\section{Wrong sign
\texorpdfstring{$h_1 b \bar{b}$}{h1bb} couplings in Type II C2HDM}
\label{sec:wrong_sign}

Recently there has been great interest in
probing the wrong sign $h b \bar{b}$ couplings,
in the context of the real 2HDM
\cite{Carmi:2012yp, Chiang:2013ixa, rui, Ferreira:2014naa, Fontes:2014tga}.
Here we discuss for the first time this issue
in the context of the Type II C2HDM.

In the Type II real 2HDM the coupling of $h_1=h$
with the down-type quarks and the charged leptons
may be written as
$m_f k_D/v$,
where $m_f$ is the mass of the appropriate
fermion,
and
\be
k_D = - \frac{\sin{\alpha}}{\cos{\beta}}\, .
\label{kD}
\ee
Here,
$\alpha$ is the angle mixing the two CP even
scalar components into a light scalar $h$
and a heavy scalar $H$.
Thus,
$\sin{\alpha}$ negative (positive) corresponds
to the (opposite of the) SM sign for $k_D$ in Type II.
Given the experimental lower bound on $\tan{\beta}$,
the coupling to the up-type quarks in Type I and Type II,
as well as the coupling to the down-type quarks
in Type I cannot have the wrong sign.
The regions of Type II with right and wrong sign
are disjoint in that the current measurements
of $\mu_{VV}$ force
$\sin{(\beta - \alpha)} \sim +1$ when $k_D>0$
and $\sin{(\beta + \alpha)} \sim +1$ when $k_D<0$
(dubbed, the wrong-sign solution).
To be precise and independent of the phase conventions
leading to the usual choices for the ranges of $\alpha$,
one should talk about $C k_D >0$ as the right sign solution
and $C k_D <0$ as the wrong sign solution,
where $C = \sin{(\beta-\alpha)}$
is the $h VV$ couplings in the real 2HDM,
divided by the $h_{\textrm{SM}} VV$ coupling in the
SM.\footnote{Rui Santos,
private communication.}

\begin{figure}[tbp]
\centering
\includegraphics[width=.45\textwidth]{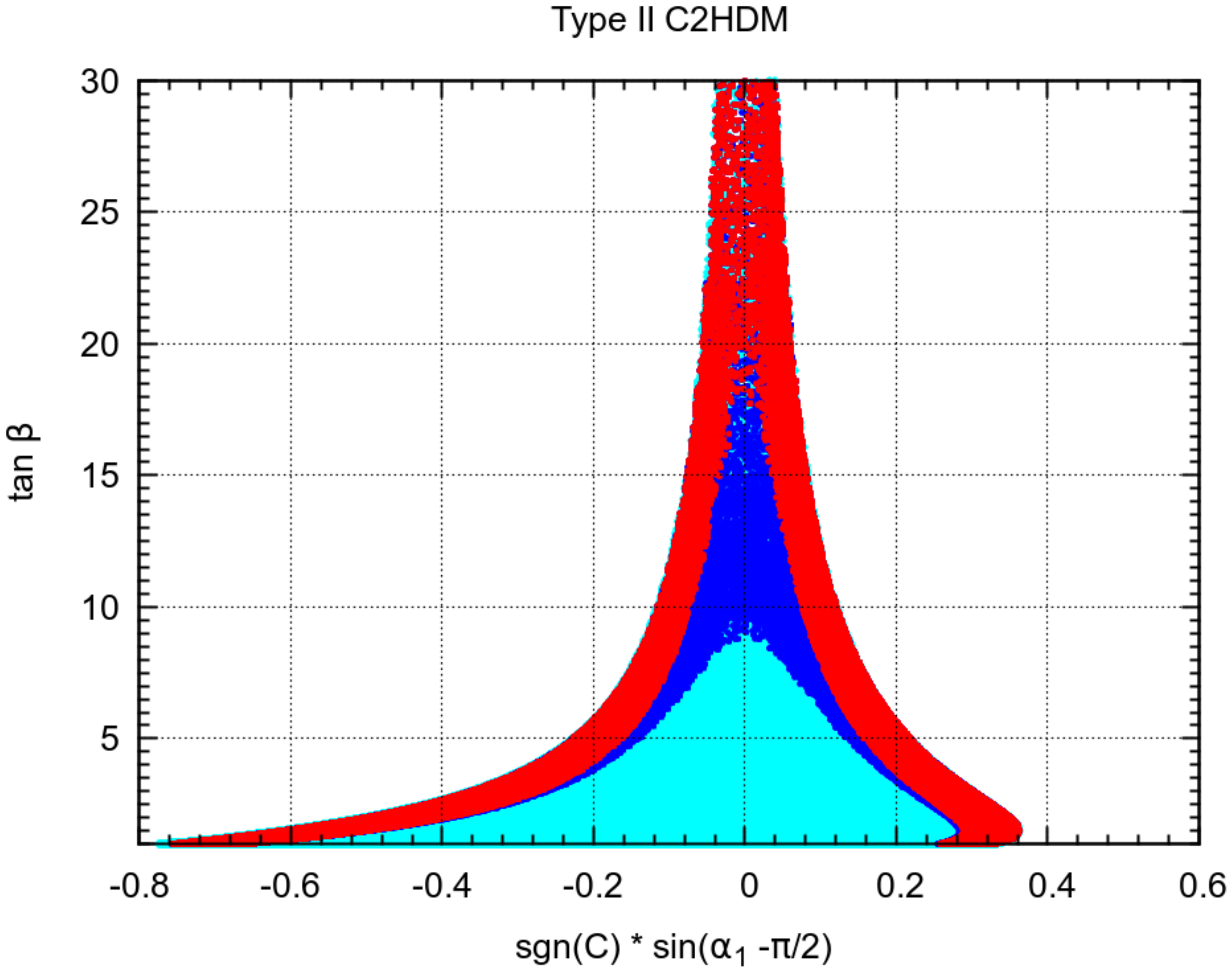}
\hfill
\includegraphics[width=.45\textwidth]{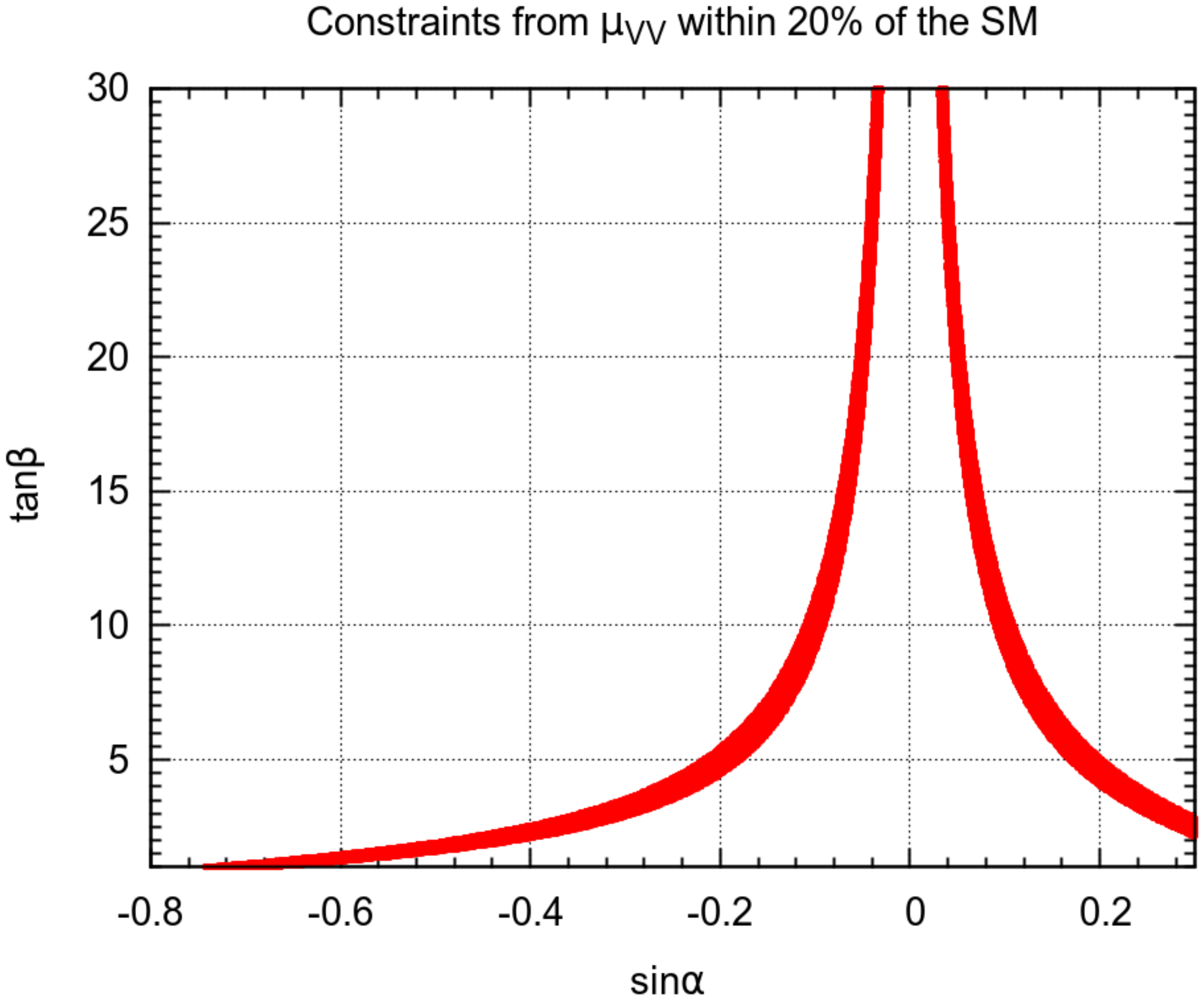}
\caption{\label{fig:sina_tanb} On the left (right)
panel,
we show the results of the simulation
of Type II C2HDM (real 2HDM) on the
$\textrm{sgn}(C)\, \sin{(\alpha_1 - \pi/2)}$-$\tan{\beta}$
($\sin{\alpha}$-$\tan{\beta}$)
plane.
On the left panel,
in cyan/light-grey we show all points
obeying $\mu_{VV} = 1.0 \pm 0.2$;
in blue/black the points that satisfy
in addition $|s_2|,\, |s_3| < 0.1$;
and in red/dark-grey the points that satisfy
$|s_2|,\, |s_3| < 0.05$.}
\end{figure}

The situation is rather different in the C2HDM because,
according to eq.~\eqref{LY},
there are two couplings of $h_1$ with the fermions:
the scalar-like coupling $a$,
and the pseudoscalar-like coupling $b$.
We follow the spirit of refs.~\cite{Ferreira:2014naa,pseudo}
and assume that experiments have obtained
the SM values for
$\mu_{ZZ}$,
$\mu_{\gamma\gamma}$,
and $\mu_{\tau^+\tau^-}$
within $20\%$.
Denoting by $\textrm{sgn}(C)$ the sign of $C$,
we show in fig.~\ref{fig:sina_tanb} a simulation in
the $\textrm{sgn}(C)\, \sin{(\alpha_1 - \pi/2)}$-$\tan{\beta}$
plane.
This reduces to the well known
$\sin{\alpha}$-$\tan{\beta}$ plane of the real 2HDM,
with the usual angle conventions,
when we take the limit $|s_2| \ra 0$ and $|s_3| \ra 0$.
In cyan/light-grey we show the points which pass $\mu_{VV} = 1.0 \pm 0.2$;
in blue/black the points that also satisfy
$|s_2|,\, |s_3| < 0.1$;
and in red/dark-grey the points that satisfy
$|s_2|,\, |s_3| < 0.05$.
The left panel of fig.~\ref{fig:sina_tanb} should be
compared with the right panel,
obtained in the real 2HDM.
The left leg of that panel corresponds to
$\sin{(\beta-\alpha)} \sim 1$ and the right sign solution,
while the right leg corresponds to  $\sin{(\beta+\alpha)} \sim 1$
and the wrong sign solution.
We see that,
for generic $s_2$ and $s_3$,
the two regions are continuously connected.
In contrast,
when $|s_2|,\, |s_3| < 0.05$,
we tend to the disjoint solutions of the real 2HDM,
as we should.

\begin{figure}[tbp]
\centering 
\includegraphics[width=.45\textwidth]{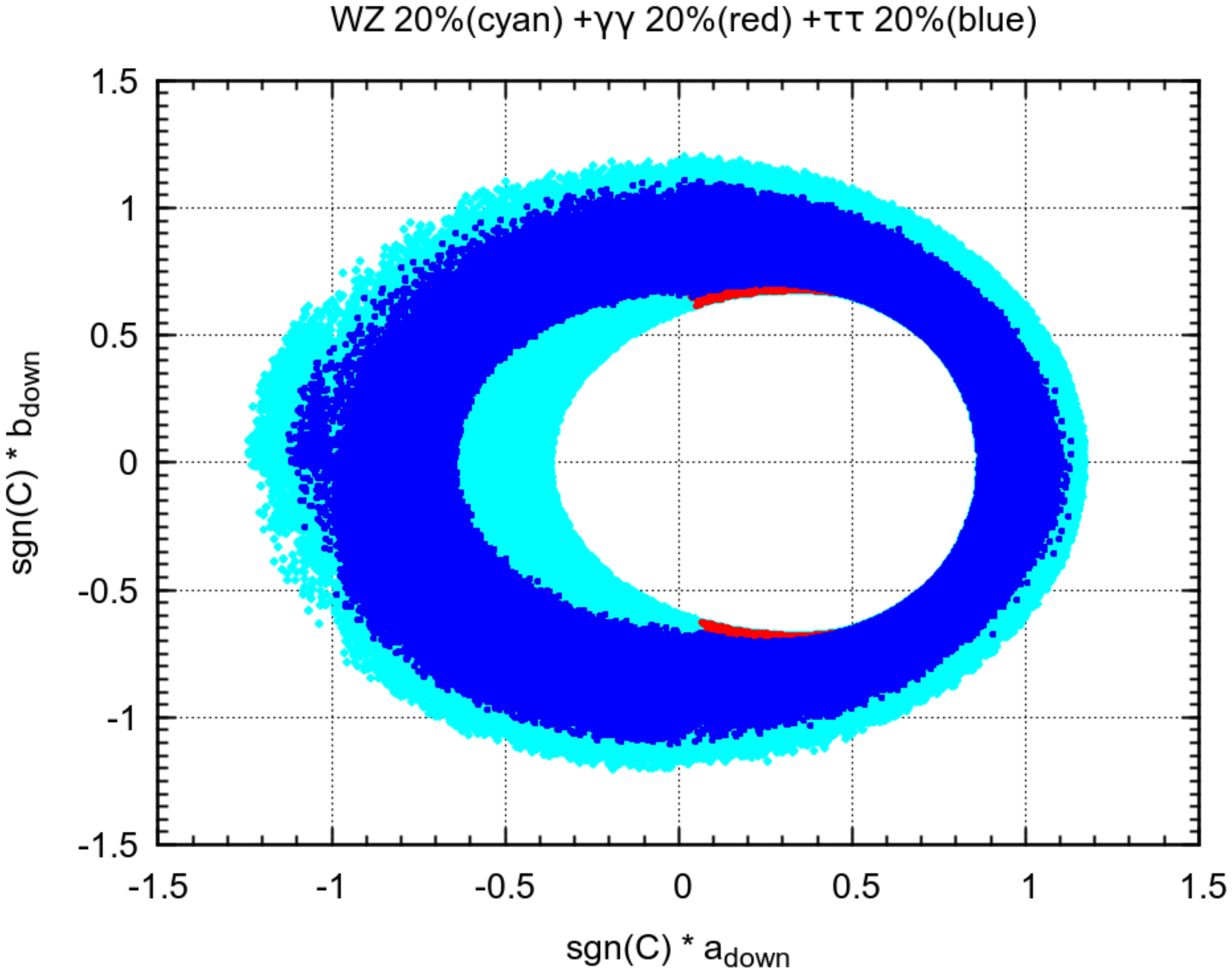}
\hfill
\includegraphics[width=.45\textwidth]{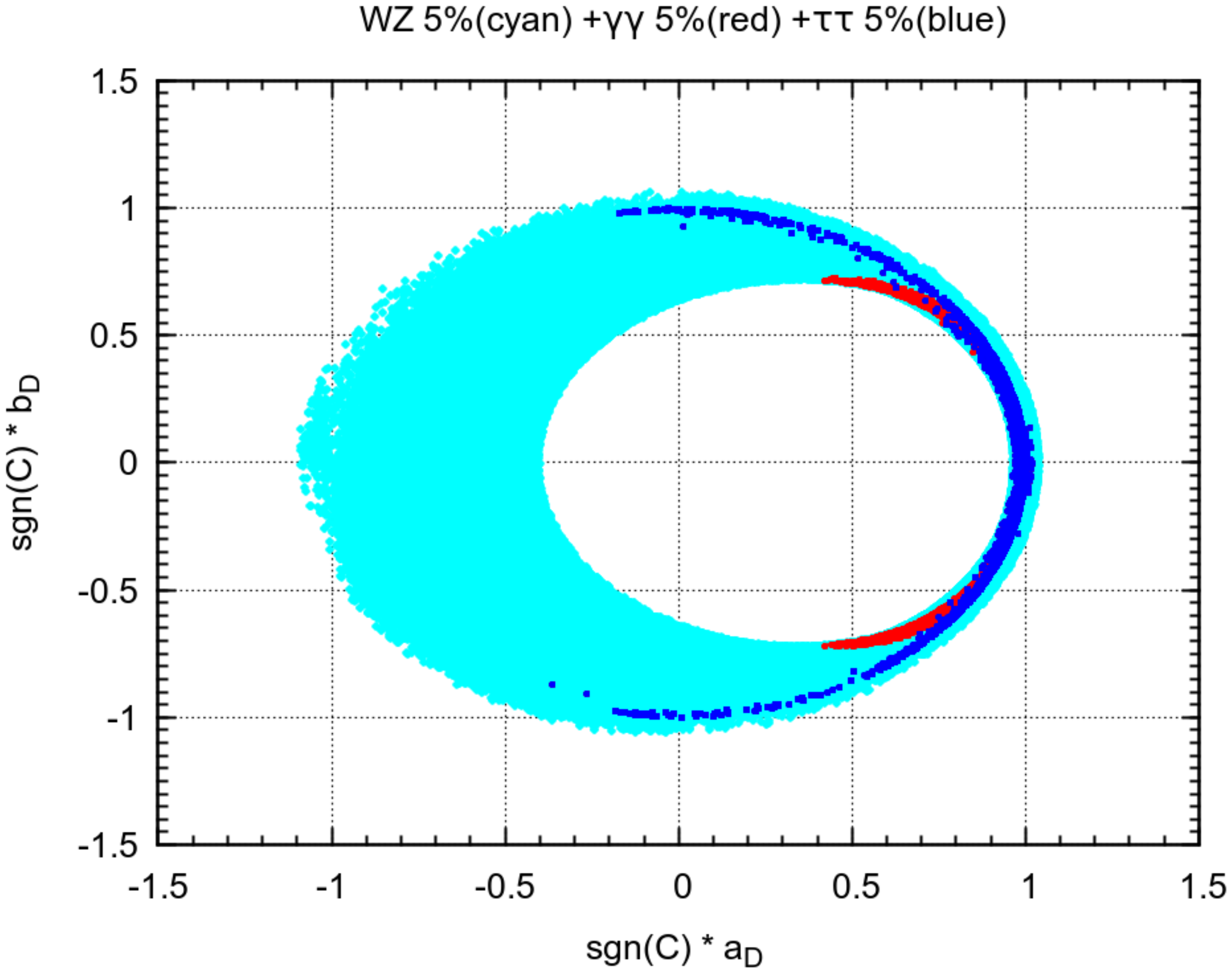}
\caption{\label{fig:aDbD_1} Results of the simulation
of Type II C2HDM on the
$\textrm{sgn}(C)\, a_D$-$\textrm{sgn}(C)\, b_D$ plane
of scalar-pseudoscalar couplings of $h_1 b \bar{b}$.
On the left panel (right panel) we assume that the measurements
come from current data at 8 TeV (prospective data at 14 TeV)
and are made
within $20\%$ ($5\%$) of the SM.
Constraints from
$\mu_{VV}$ are in cyan/light-grey,
from
$\mu_{\gamma\gamma}$ are in red/dark-grey,
and from $\mu_{\tau^+\tau^-}$ are in blue/black.}
\end{figure}

The constraints on the
$\textrm{sgn}(C)\, a_D$-$\textrm{sgn}(C)\, b_D$ plane
are shown on the left panel of fig.~\ref{fig:aDbD_1}.
We see that $\textrm{sgn}(C)\, a_D$ can have both signs
(as it could in the CP conserving limit,
where $a_D=k_D$),
and so can $\textrm{sgn}(C)\, b_D$.
Moreover, these different regions are continuously connected.
In the C2HDM there is still a very large region of either
negative sign permitted.
The situation will be altered if future measurements
fix $\mu_{VV}$,
$\mu_{\gamma\gamma}$,
and $\mu_{\tau^+\tau^-}$
to within $5\%$ of the SM,
as shown on the right panel of fig.~\ref{fig:aDbD_1}.
In that case,
there will be almost no region with
$\textrm{sgn}(C)\, a_D<0$.
This is consistent with the disappearance of the
negative $k_D$ region in the real Type II 2HDM
when the measurements reach the $5\%$ level \cite{Ferreira:2014naa}.
However,
in the C2HDM some points with $\textrm{sgn}(C)\, a_D \sim -0.4$
are allowed,
if one also has a large pseudoscalar coupling
$\textrm{sgn}(C)\, b_D \sim -0.8$.

\begin{figure}[tbp]
\centering 
\includegraphics[width=.45\textwidth]{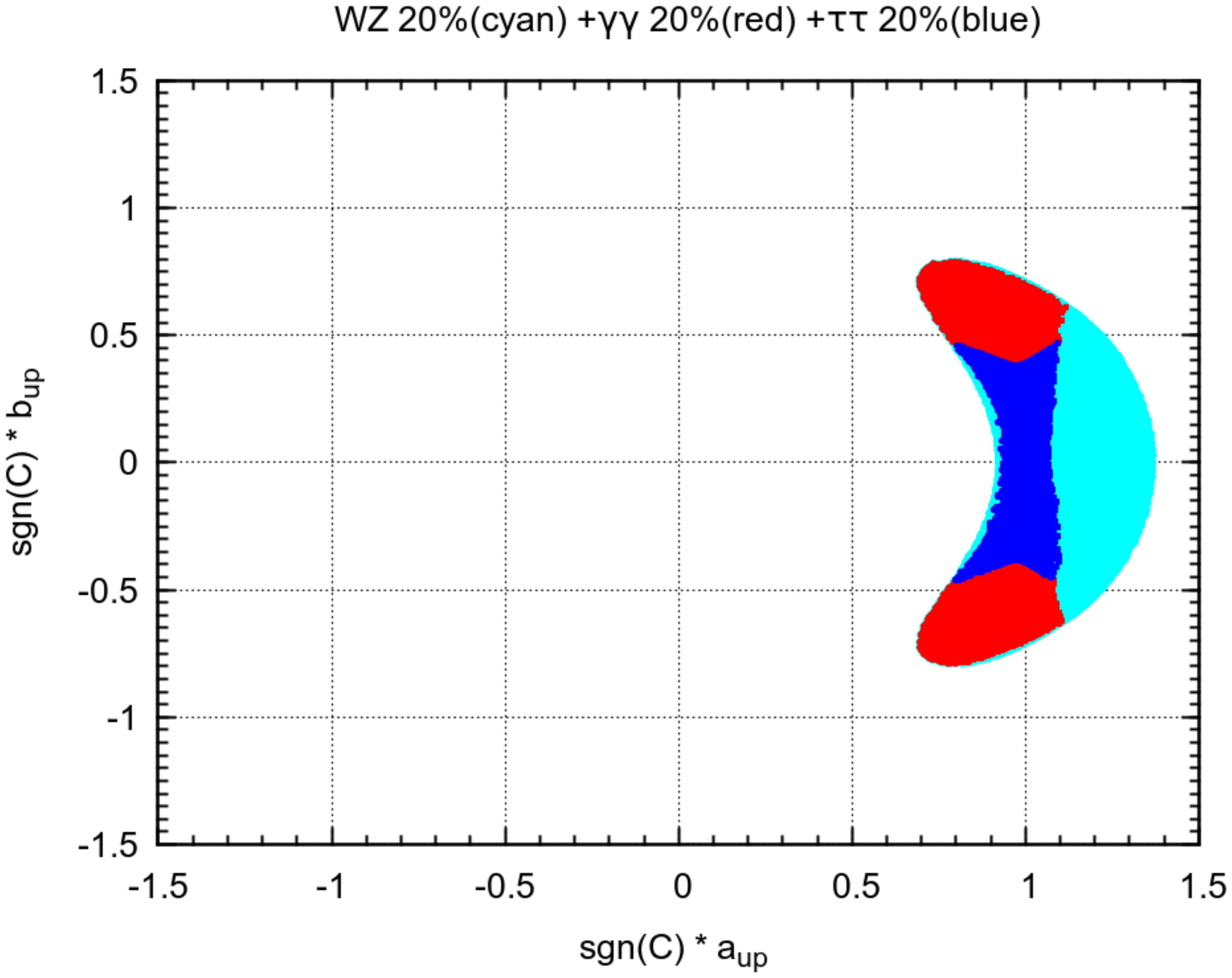}
\hfill
\includegraphics[width=.45\textwidth]{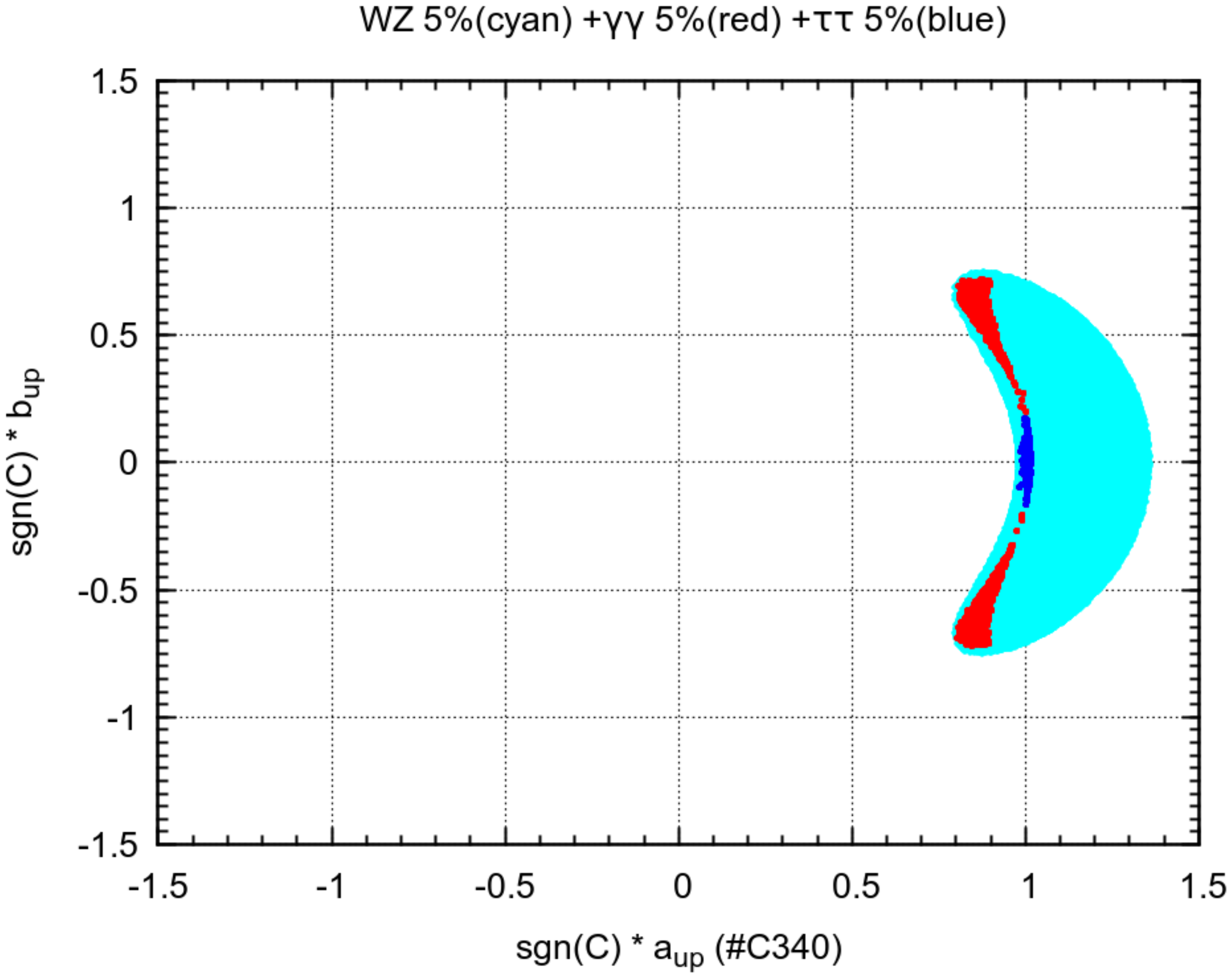}
\caption{\label{fig:aUbU_1} Results of the simulation
of Type II C2HDM on the
$\textrm{sgn}(C)\, a_U$-$\textrm{sgn}(C)\, b_U$ plane
of scalar-pseudoscalar couplings of $h_1 t \bar{t}$.
On the left panel (right panel) we assume that the measurements
come from current data at 8 TeV (prospective data at 14 TeV)
and are made within $20\%$ ($5\%$) of the SM.
Constraints from
$\mu_{VV}$ are in cyan/light-grey,
from
$\mu_{\gamma\gamma}$ are in red/dark-grey,
and from $\mu_{\tau^+\tau^-}$ are in blue/black.}
\end{figure}

In the real 2HDM,
the lower bound $\tan{\beta} > 1$ implies that
the coupling of $h t \bar{t}$ must be positive.
In the C2HDM,
it is still true that the scalar like coupling $\textrm{sgn}(C)\, a_U$
must be positive,
but the pseudoscalar like $\textrm{sgn}(C)\, b_U$ can have either sign.
This is illustrated in
fig.~\ref{fig:aUbU_1},
for measurements within $20\%$ (left panel)
and $5\%$ (right panel) of the SM.
Notice that $\mu_{\gamma\gamma}$ forces the figure into the
outer rim, and that adding $\mu_{\tau^+\tau^-}$
forces $\textrm{sgn}(C)\, a_U \sim 1$ and $|b_U| \lesssim 0.2$.
This shows that the line of blue/black points which one
guesses on the right panel of fig.~\ref{fig:aDbD_1}
corresponds to $\textrm{sgn}(C)\, a_U \sim 1$.

\begin{figure}[tbp]
\centering 
\includegraphics[width=.45\textwidth]{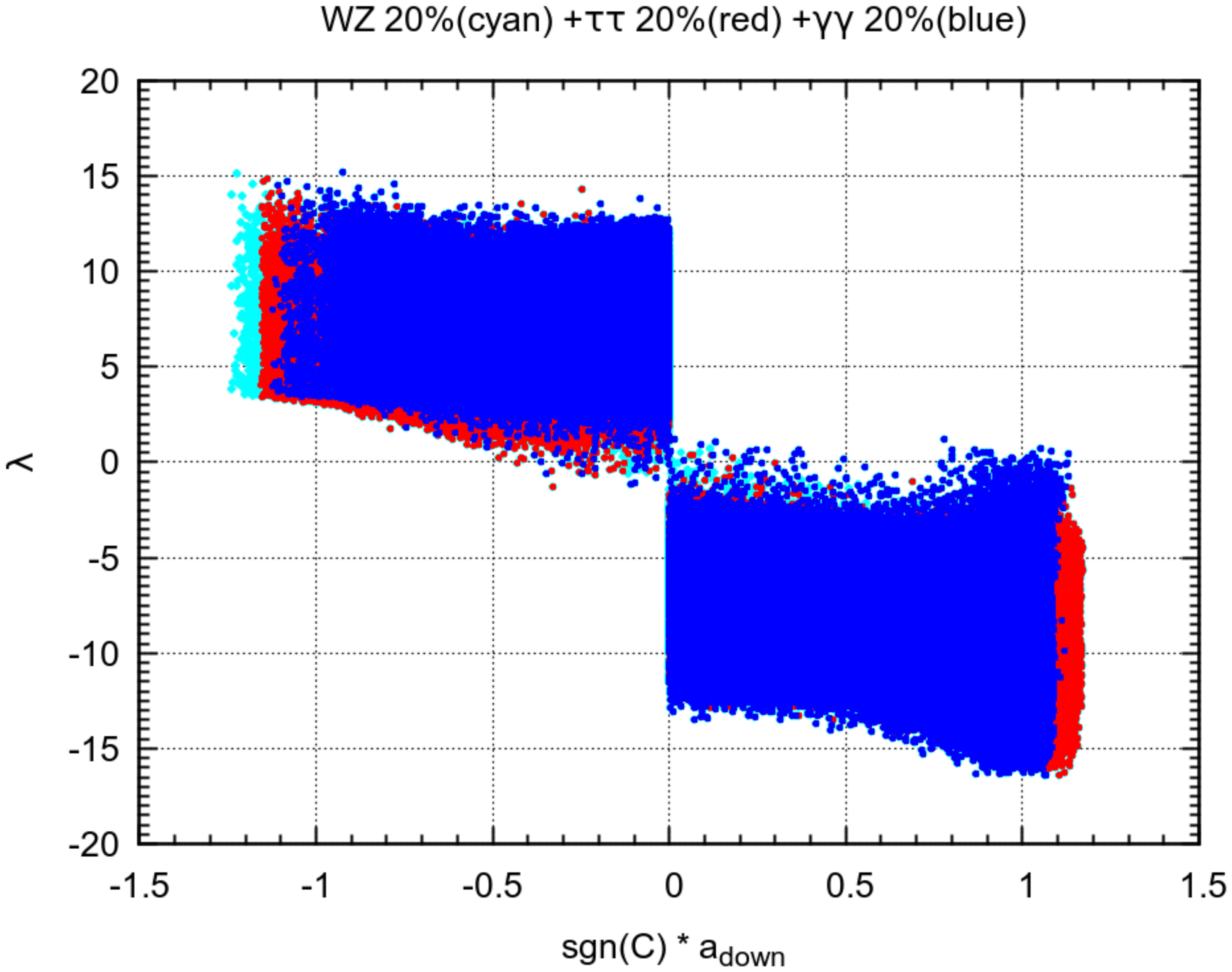}
\hfill
\includegraphics[width=.45\textwidth]{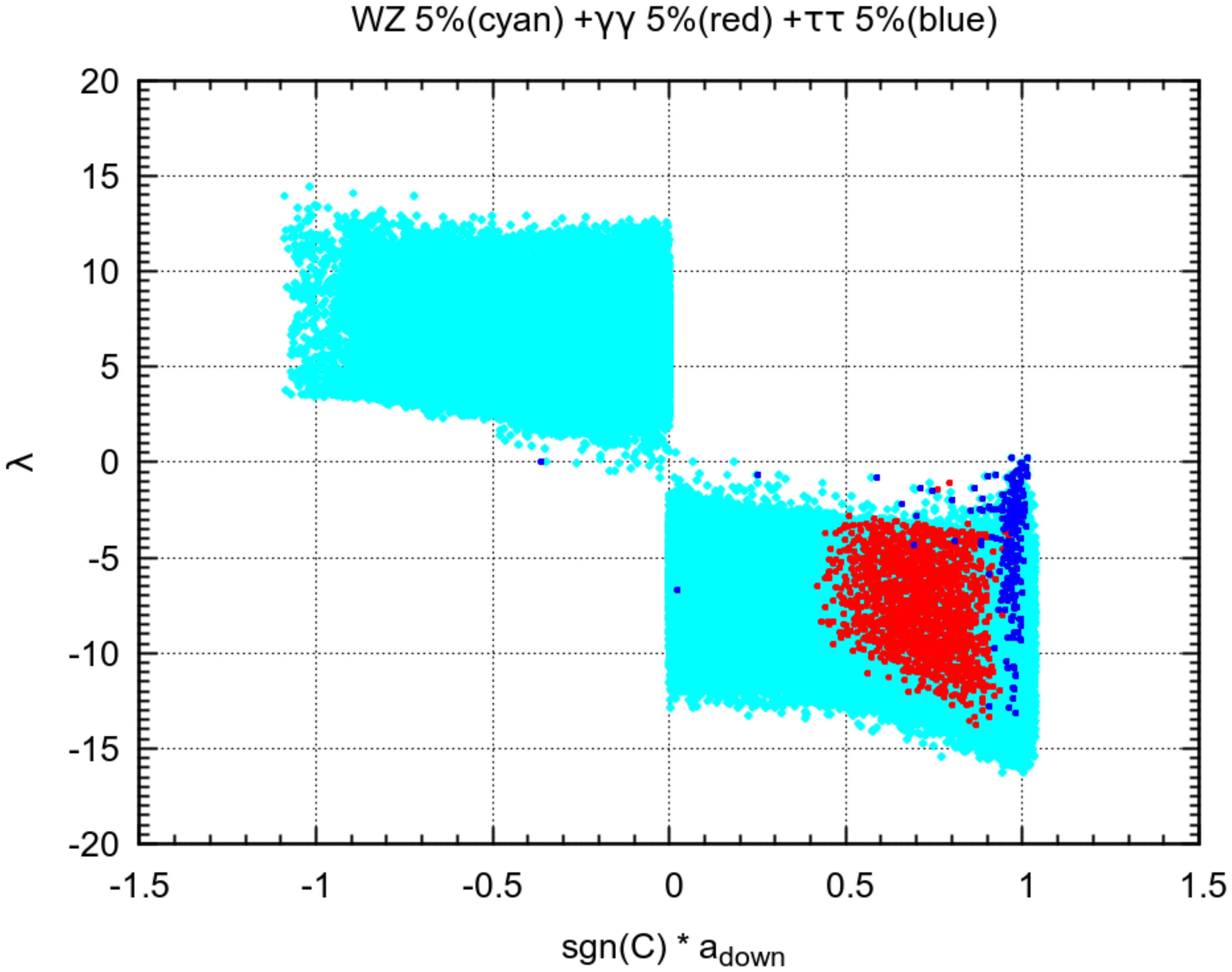}
\caption{\label{fig:lambda} Results of the simulation
of Type II C2HDM on the
$\textrm{sign}(C)\, a_D$-$\lambda$ plane.
On the left panel (right panel) we assume that the measurements
come from current data at 8 TeV (prospective data at 14 TeV)
and are made
within $20\%$ ($5\%$) of the SM.
Constraints from
$\mu_{VV}$ are in cyan/light-grey;
adding constraints from
$\mu_{\tau^+\tau^-}$
($\mu_{\gamma\gamma}$ at $5\%$)
only the points in
red/dar-grey survive;
adding constraints from
$\mu_{\gamma\gamma}$
($\mu_{\tau^+\tau^-}$ at $5\%$)
only the
points in blue/black,
survive.}
\end{figure}

A final point of interest concerns the effect on delayed decoupling.
In the real 2HDM,
wrong sign solutions exist only with $k_D \sim -1$.
In fact,
as explained in \cite{Fontes:2014tga},
a rather simple trigonometric explanation
justifies that a 20\% bound on $\mu_{VV}$ implies an even better
determination of $\sin^2{(\beta-\alpha)}$ for
a given $\tan{\beta}$.\footnote{For example,
for $\tan{\beta} = 10$,
a 20\% bound around $\mu_{VV} \sim 1$  implies
a determination of $\sin^2{(\beta-\alpha)}$ to better than
$0.5\%$.}
As pointed out in ref.~\cite{Ferreira:2014naa},
this solution exists if and only if the charged Higgs loop gives
a contribution of order $10\%$ to $h \ra \gamma\gamma$,
due to the fact that the $h H^+ H^-$ coupling $\lambda$
-- see Eq.~\eqref{LhHpHm} --
exhibits a non-decoupling with the charged Higgs mass,
curtailed only by the requirements of unitarity.
In fig.~\ref{fig:lambda},
we show what happens to $\lambda$
as a function of
$a_D$ multiplied by the sign of $C$.
On the left panel of fig.~\ref{fig:lambda},
the points in cyan/light-grey pass $\mu_{VV} = 1$
within $20\%$.
The points in red/dark-grey pass
this constraint and, in addition,
$\mu_{\tau^+\tau^-} = 1$
within $20\%$.
The points in blue/black pass
the previous two constraints and, in addition,
$\mu_{\gamma\gamma} = 1$
within $20\%$.
These simulations were made at 8 TeV to allow a feeling for the
current constraints.
The colour code on the right panel are:
cyan/light-grey points pass $\mu_{VV} = 1$
within $5\%$;
red/dark-grey points pass in addition
$\mu_{\gamma\gamma} = 1$
within $5\%$;
blue/black points pass
the previous two constraints and, in addition,
$\mu_{\gamma\gamma} = 1$
within $5\%$.
These prospective simulations have been drawn at 14 TeV.

From the left panel of fig.~\ref{fig:lambda},
we see that in the C2HDM one can have
any value for $\textrm{sign}(C)\, a_D$
between around $-1.1$ and $1.05$.
This is different from the real 2HDM where
$k_D \sim 1$ and $k_D \sim -1$ form two disjoint solutions.
The difference, of course,
is due to the fact that in the C2HDM there is a new
pseudoscalar coupling $b_D$.
But there is a similarity.
Indeed, values of $\textrm{sign}(C)\, a_D \sim -1$
correspond to non-negligible values for $\lambda$,
as seen on the left panel of fig.~\ref{fig:lambda}.
This is the analogous of the delayed decoupling
found for $k_D \sim -1$ solutions found in the real 2HDM.
The right panel of fig.~\ref{fig:lambda}
shows again that a putative $5\%$ future measurement around
the SM to be made at 14 TeV will eliminate almost all the
$\textrm{sign}(C)\, a_D < 0$ points.

\section{Conclusions}
\label{sec:conclusions}

The 125 GeV particle found at LHC could have a pseudoscalar
component.
We discuss in detail the decay of a mixed scalar/pseudoscalar state
into $Z \gamma$,
which will be probed in the next LHC run.
We consider the constraints that current experiments impose
on the four versions of the C2HDM and discuss the prospects
of future bounds, including $h \ra Z \gamma$.
This provides an update of Type I and Type II,
and the first discussion of current constraints on
the Lepton Specific and Flipped C2HDM.

In the C2HDM,
the parameter $s_2$ measures the pseudoscalar content,
with $s_2=0$ ($|s_2|=1$) corresponding to a pure
scalar (pseudoscalar).
The fact that ATLAS has a rather large central value
for $\mu_{\gamma\gamma}$ places strong limits on C2HDM,
but it also disfavours the SM at 2-$\sigma$.
But,
even excluding this constraint,
we find that current experiments already disfavor
a large pseudoscalar component $|s_2|>0.85$,
at over 1-$\sigma$ level in all C2HDM versions.

As for future experimental reaches,
we find that in all types of C2HDM a better measurement
of $\mu_{b \bar{b}}(Vh) \sim 1$ will exclude large values
of the pseudoscalar component $s_2$.
Similarly,
a measurement  of $\mu_{Z \gamma} \sim 1$ will
also exclude a very large $s_2$ component.
The Flipped C2HDM is special in that one can have
$\mu_{\tau^+ \tau^-} \sim 7$ and,
thus,
the $\tau^+ \tau^-$ channel will be crucial in probing
this model.

Finally,
we have discussed the possibility that the
scalar component of the Type II C2HDM
$h_1 q \bar{q}$  coupling ($a$)
has a sign opposite to that in the SM.
The fact that the C2HDM also has
a pseudoscalar component of the $h_1 q \bar{q}$ coupling ($b$)
gives more room for differences than are possible within the
Type II real 2HDM.
We found that the up quark coupling
$\textrm{sgn}(C)\, b_U$ can have either sign,
while $\textrm{sgn}(C)\, a_U$ must be positive.
If future experiments yield
$\mu_{VV}$,
$\mu_{\gamma\gamma}$,
and $\mu_{\tau^+ \tau^-}$
within $5\%$ of the SM,
then $\textrm{sgn}(C)\, b_U$ can still
have either sign,
but $\textrm{sgn}(C)\, a_U=1$ to very high
precision,
corresponding to the limit $s_1 c_2 = s_\beta$.
In contrast,
current experiments allow
for either sign of both
$\textrm{sgn}(C)\, a_D$
and
$\textrm{sgn}(C)\, b_D$,
covering a rather large region.
However,
if future experiments yield
$\mu_{VV}$,
$\mu_{\gamma\gamma}$,
and $\mu_{\tau^+ \tau^-}$
within $5\%$ of the SM,
then the region in the
$\textrm{sgn}(C)\, a_D$-$\textrm{sgn}(C)\, b_D$ plane
reduces to a line, with most points concentrated around
$\textrm{sgn}(C)\, a_D \sim 1$.
Still,
there are a few points with $\textrm{sgn}(C)\, a_D \sim -0.4$,
as long as $\textrm{sgn}(C)\, b_D \sim -0.8$ is rather
large.

\appendix

\section{Production and Decay rates}
\label{app:rates}

\subsection{Lagrangian}

The appendices contains the production and decay rates for
a scalar particle with both scalar and pseudo-scalar components.
We assume that the SM particles except the Higgs follow the usual
lagrangian, that there are $H^\pm$ particles with the usual
gauge-kinetic Lagrangian,
and that the new scalar/pseudoscalar
particle $h$ has the following interactions:
\ba
{\cal L}_Y
&=&
-\left(\sqrt{2} G_\mu \right)^{\tfrac{1}{2}}\, m_f\
\bar \psi \left( a + i b \gamma_5 \right) \psi\, h,
\label{LY}
\\
{\cal L}_{h H^+ H^-}
&=&
\lambda\, v\, h H^+ H^-,
\label{LhHpHm}
\\
{\cal L}_{h V V}
&=&
C \left[
g\, m_W W_\mu^+ W^{\mu -} + \frac{g}{2 c_W} m_Z Z_\mu Z^\mu
\right]\, h,
\label{LVV}
\ea
where $a$, $b$, and $C$ are real, $c_W = \cos{\theta_W}$,
and $\theta_W$ is the Weinberg angle.
In the limit, $a=C=1$, and $b = \lambda=0$,
we obtain the SM.

We use the notation for the covariant derivatives
contained in Rom\~{a}o and Silva \cite{Romao:2012pq},
with all etas positive, which coincides with the convention in \cite{HHG}.
Some relevant vertices are
\ba
h \bar{\psi} \psi
& \ \ \rightarrow\ \ &
- i\, \frac{g\, m_f}{2 m_W} \left( a + i b \gamma_5 \right)\, ,
\nonumber\\[+2mm]
h H^+ H^-
& \ \ \rightarrow\ \ &
i\, \lambda\, v\, ,
\nonumber\\[+2mm]
h W^{+ \mu} W^{- \nu}
& \ \ \rightarrow\ \ &
i\, g\, m_W\, C\, g^{\mu \nu}\, ,
\nonumber\\[+2mm]
h Z^\mu Z^\nu
& \ \ \rightarrow\ \ &
i\, \frac{g\, m_Z}{\cos(\theta_W)}\, C\, g^{\mu \nu}\, ,
\nonumber\\[+2mm]
H^+ H^- A^\mu
& \ \ \rightarrow\ \ &
- i e\, (p_+  -   p_-)^\mu\, ,
\nonumber\\[+2mm]
H^+ H^- Z^\mu
& \ \ \rightarrow\ \ &
- i g\, \frac{\cos(2\theta_W)}{2 \cos(\theta_W)}
(p_+  - p_-)^\mu\, ,
\nonumber\\[+2mm]
H^+ H^- A^\mu A^\nu
& \ \ \rightarrow\ \ &
2 i\, e^2  g^{\mu\nu}\, ,
\nonumber\\[+2mm]
H^+ H^- Z^\mu A^\nu
& \ \ \rightarrow\ \ &
i e g\, \frac{\cos(2\theta_W)}{\cos(\theta_W)}
g^{\mu\nu}\, .
 \label{eq:22}
\ea
These couplings were checked for the 2HDM with FeynRules \cite{feynrules}
with the conventions of Rom\~{a}o and Silva \cite{Romao:2012pq} for positive
$\eta$s.

\subsection{Tree level production and decay}

In this article,
we use
\be
\tau =
4 m^2/m_h^2,
\label{tau}
\ee
where $m$ is the mass of the relevant particle
while $m_h=125$ GeV.
This is the notation of \cite{HHG}.
In \cite{pseudo,djouadi1,djouadi2}
the notation is $\tau (\textrm{theirs})=\tau^{-1}$.

The decays into fermions are given by
\be
\Gamma(h \rightarrow f \bar{f})
=
N_c \frac{G_\mu\, m_f^2}{4 \sqrt{2} \pi} m_h
\left[ a^2 \beta_f^3 + b^2 \beta_f \right],
\label{hffdecay}
\ee
where $N_c=3$ ($N_c=1$) for quarks (leptons)
and
$\beta_f = \sqrt{1 - 4 m_f^2/m_h^2} = \sqrt{1 - \tau}$.
The decays into two vector bosons are given by
\be
\Gamma(h \rightarrow V^{(\ast)} V^{(\ast)})
=
C^2\
\Gamma_{\textrm{SM}}(h \rightarrow V^{(\ast)} V^{(\ast)}),
\ee
and the partial decay widths in the SM-Higgs case in the two-, three-
and four-body approximations,
$\Gamma_{\textrm{SM}}(h \rightarrow V^{(\ast)} V^{(\ast)})$,
can be found in Section I.2.2 of ref.~\cite{djouadi1}.

For the vector boson fusion (VBF)
and associated (VH) productions, we find
\be
\frac{\sigma_{\textrm{VBF}}}{\sigma^{\textrm{SM}}_{\textrm{VBF}}}
=
\frac{\sigma_{\textrm{VH}}}{\sigma^{\textrm{SM}}_{\textrm{VH}}}
= C^2,
\ee
while, for the $b \bar{b}$ production,
\be
\frac{\sigma (b \bar{b} \rightarrow h)}{
\sigma^{\textrm{SM}} (b \bar{b}  \rightarrow h)}
= a^2 + b^2.
\label{bbprod}
\ee

We point out that the expressions shown here hold for any model
with the effective Lagrangians of Eqs.~\eqref{LY}-\eqref{LVV}.
Also,
there is no interference between the scalar $a$ couplings
and the pseudoscalar $b$ couplings in Eqs.~\eqref{hffdecay} or \eqref{bbprod}.

\section{Amplitudes for
\texorpdfstring{$h \ra \gamma \gamma$}{h --> gamma gamma}
}

\subsection{Fermion Loop}

The relevant interaction for the fermion loop is in \eqref{LY}.
The one-loop amplitude reads
\begin{equation}
M_F^{\gamma\gamma} \equiv
\left(
q_1 \cdot q_2\, \epsilon_1 \cdot \epsilon_2  -
q_1 \cdot  \epsilon_2\, q_2 \cdot  \epsilon_1
\right)
c_{F}^{\gamma\gamma}
+
\epsilon_{\mu\nu\alpha\beta} \ q_1^\mu q_2^\nu \epsilon_1^\alpha
\epsilon_2^\beta \ d_{F}^{\gamma\gamma}\, ,
\label{eq:2}
\end{equation}
where
\ba
c_F^{\gamma\gamma}
&=&
- \frac{e^2 Q_f^2 g}{m_W}
\frac{4 a\, m_f^2}{m_h^2} \frac{1}{16\pi^2}
\left[
\left( 4  m_f^2 - m_h^2 \right)
C_0(0,0,m_h^2, m_f^2,m_f^2,m_f^2) +  2
\right]\, ,
\nonumber\\[+2mm]
d_F^{\gamma\gamma}
&=&
4 \frac{e^2 Q_f^2 g}{m_W} \frac{1}{16\pi^2} b\,
m_f^2\ C_0(0,0,m_h^2, m_f^2,m_f^2,m_f^2)\, ,
\ea
where $C_0$ is one of the Passarino-Veltman \cite{passarino:1979jh}
scalar loop integrals.
Their relation with other expressions for the one loop
integrals is explained in appendix~\ref{app:relation}.
Note that the definition of the amplitude in eq.~\eqref{eq:2} is the
same as in ref.~\cite{HHG},
but differs by an irrelevant global
sign from the definition in refs.~\cite{pseudo,djouadi1}.

To make contact with the more conventional notation we define
\begin{equation}
c_F^{\gamma\gamma}\equiv \frac{e^2 g}{m_W}\frac{1}{16\pi^2}
X_F^{\gamma\gamma},
\quad
d_F^{\gamma\gamma}\equiv \frac{e^2g}{
m_W}\frac{1}{16\pi^2} Y_F^{\gamma\gamma},
\quad
\tau_f \equiv \frac{4 m_f^2}{m_h^2}\, .
\label{eq:3}
\end{equation}
We then get
\ba
X_F^{\gamma\gamma}
&=&
- \frac{4 a\, Q_f^2\, m_f^2}{m_h^2}
\left[
\left( 4 m_f^2 - m_h^2 \right)
C_0(0,0,m_h^2, m_f^2,m_f^2,m_f^2) + 2
\right]
\nonumber\\[+2mm]
Y_F^{\gamma\gamma}
&=&
4 b\, Q_f^2\, m_f^2\ C_0(0,0,m_h^2, m_f^2,m_f^2,m_f^2)
\ea
Finally, using
\begin{equation}
C_0(0,0,m_h^2, m_f^2,m_f^2,m_f^2) = - \frac{\tau_f f(\tau_f)}{2 m_f^2},
\label{eq:4}
\end{equation}
where $f(\tau)$ is the function defined in the
Higgs Hunter's Guide \cite{HHG},
\begin{equation}
f(\tau)=
\left\{
\begin{array}{ll}
\left[\sin^{-1} \left(\sqrt{1/\tau}\right)\right]^2,
&\ \  \text{if}\ \tau\ge 1\ ,\\[+2mm]
-\frac{1}{4}
\left[ \displaystyle
\ln\left(\frac{1+\sqrt{1-\tau}}{1-\sqrt{1-\tau}}\right)
-i \pi \right]^2 ,
&\ \  \text{if}\ \tau<1\, ,
\end{array}
\right.
\label{eq:9}
\end{equation}
we obtain (summing over all fermions)
\ba
X_F^{\gamma\gamma}
&=&
- \sum_f N_c^f 2 a\, Q_f^2\, \tau_f \left[ 1 + (1-\tau_f)
f(\tau_f)\right]\, ,
\nonumber\\[+2mm]
Y_F^{\gamma\gamma}
&=&
-\sum_f N_c^f  2 b\, Q_f^2\, \tau_f f(\tau_f)\, .
\ea

\subsection{Gauge boson loops}

As the only modification introduced by the new Lagrangian is
a multiplicative constant $C$, we can use the SM
result ($C=1$ in the SM).
Using the same notation as in
eq.~\eqref{eq:3},
we get~\cite{HHG},
\begin{equation}
X_W^{\gamma\gamma}
=
C \left[\vb{12} 2 + 3 \tau_W + 3 \tau_W (2-\tau_W)
f(\tau_W)\right]\, ,
\end{equation}
and, of course, $Y_W^{\gamma\gamma}=0$.

\subsection{Charged Higgs loops}

We get for the three diagrams contributing to this process,
\begin{equation}
M_H^{\gamma\gamma} =
\left(
q_1 \cdot q_2\, \epsilon_1 \cdot \epsilon_2  -
q_1 \cdot  \epsilon_2\, q_2 \cdot  \epsilon_1
\right)
c_{H}^{\gamma\gamma}\, ,
\label{eq:10}
\end{equation}
where
\begin{equation}
c_{H}^{\gamma\gamma} =
- \frac{4 e^2 \lambda v}{m_h^2 16\pi^2}
\left[
\vb{12} 2 m_{H^\pm}^2
C_0(0,0,m_h^2,m_{H^\pm}^2,m_{H^\pm}^2,m_{H^\pm}^2)
+ 1
\right]\, .
\label{eq:11First}
\end{equation}
In the notation of eq.~\eqref{eq:3} we get
\begin{align}
  \label{eq:12}
  X_H^{\gamma\gamma} =& - \frac{4\lambda m_W v}{g m_h^2} \left[\vb{12}
2 m_{H^\pm}^2
C_0(0,0,m_h^2,m_{H^\pm}^2,m_{H^\pm}^2,m_{H^\pm}^2)+1\right]
\nonumber \\[+2mm]
=& - \frac{\lambda v^2}{2 m_{H^\pm}^2} \tau_\pm \left[\vb{11}
  1-\tau_\pm f(\tau_\pm) \right]\, .
\end{align}
Note that this is in agreement with eq.~(2.17) of ref.~\cite{HHG},
despite the apparent sign difference, because our definition of the
coupling, in eq.~\eqref{LhHpHm},
also differs in sign from their
eq.~(2.15).
So we are in complete agreement with ref.~\cite{HHG}.
With respect to ref.~\cite{pseudo},
if we compare with their eqs.~(A.8) and (A.4),
again we differ by a global sign and we are, therefore, in agreement.
The same holds for ref.~\cite{djouadi1}.

\subsection{Renormalization and gauge invariance}

As is well known,
the loop contributions to $h \rightarrow \gamma\gamma$
should be finite and gauge invariant.
This is not achieved on a diagram by diagram basis,
but, rather,
this should be true after adding all the diagrams.
With the help of \texttt{FeynCalc}~\cite{feyncalc:1992},
we have explicitly verified this.

\section{Amplitudes for
\texorpdfstring{$h \ra Z \gamma$}{h --> Z gamma}
}
\label{app:hZgamma}

\subsection{Fermion Loop}

With the kinematics $h(p) \ra Z(q_2) \gamma(q_1)$,
the fermion loop yields an expression similar to the one
for $h \ra \gamma\gamma$:
\begin{equation}
M_F^{Z \gamma}
=
\left(
q_1 \cdot q_2\, \epsilon_1 \cdot \epsilon_2  -
q_1 \cdot  \epsilon_2\, q_2 \cdot  \epsilon_1
\right)
c^{Z \gamma}
+
\epsilon_{\mu\nu\alpha\beta}
\ q_1^\mu q_2^\nu \epsilon_1^\alpha
\epsilon_2^\beta \ d^{Z \gamma}\, .
\label{eq:2a}
\end{equation}
Again,
defining
\begin{equation}
c^{Z \gamma}\equiv \frac{e^2 g}{m_W}\frac{1}{16\pi^2} X_F^{Z \gamma}, \quad
d^{Z \gamma}\equiv \frac{e^2 g}{m_W}\frac{1}{16\pi^2} Y_F^{Z \gamma}\, ,
\label{eq:3a}
\end{equation}
we get (summing over all the fermions)
\ba
X_F^{Z \gamma}
&=&
- \sum_f N_c^f \frac{4 a\, g_V^f\, Q_f\, m_f^2}{s_W c_W}
\left[
\frac{2 m_Z^2}{(m_h^2-m_Z^2)^2}
\left[\vb{11} B_0(m_h^2,m_f^2,mf^2) -B_0(m_Z^2,m_f^2,m_f^2) \right]
\right.
\nonumber\\[+2mm]
&&
\left.
+ \frac{1}{m_h^2-m_Z^2} \left[\vb{11}
\left(4 m_f^2 - m_h^2 + m_Z^2 \right)
C_0(m_Z^2,0,m_h^2,m_f^2,m_f^2,m_f^2) +2 \right] \right]\, ,
\label{eq:15a}
\\[+2mm]
Y_F^{Z \gamma}
&=&
\sum_f N_c^f \frac{4 b\, g_V^f\, Q_f\, m_f^2}{s_W c_W}
C_0(m_Z^2,0,m_h^2, m_f^2,m_f^2,m_f^2)\, .
\label{eq:15b}
\ea

\subsection{Gauge boson loops}

As the only modification introduced by the new Lagrangian is
a multiplicative constant $C$, we can use the SM
result ($C=1$ in the SM).
Using the same notation as in eq.~\eqref{eq:3},
we get
\begin{equation}
X_W^{Z \gamma}  = \frac{C}{\tan\theta_W} I_W\, ,
\end{equation}
where
\ba
I_W
&=&
\frac{1}{(m_h^2 - m_Z^2)^2}
\left[
\vb{13}
m_h^2(1-\tan^2\theta_W)
-2m_W^2 (-5 + \tan^2\theta_W)
\right]
m_Z^2 \Delta B_0
\nonumber\\[+2mm]
&&
+\frac{1}{m_h^2 - m_Z^2}
\left[
\vb{13}
m_h^2(1-\tan^2\theta_W) -2m_W^2 (-5 +
\tan^2\theta_W)
\right.
\nonumber\\[+2mm]
&&
\left.
\vb{13}
\hspace{5mm}
+ 2 m_W^2
\left[
\vb{11}
(-5 + \tan^2\theta_W)(m_h^2 -2 M_W^2)
-2 m_Z^2 (-3 +\tan^2\theta_W)\right]
C_0
\right]\, ,
\label{eq:16}
\ea
with
\ba
\Delta B_0
&=&
B_0(m_h^2, m_W^2, m_W^2)-B_0(m_Z^2,m_W^2,m_W^2),
\nonumber\\[+2mm]
C_0
&=&
C_0(m_Z^2, 0, m_h^2, m_W^2, m_W^2, m_W^2)\, ,
\ea
and, of course, $Y_W^{Z \gamma}=0$.

\subsection{Charged Higgs loops}

There are three diagram contributions to this process.
Adding them, we get
\begin{equation}
M_{H^\pm}^{Z \gamma}
=
\left(
q_1 \cdot q_2\, \epsilon_1 \cdot \epsilon_2  -
q_1 \cdot  \epsilon_2\, q_2 \cdot  \epsilon_1
\right)
c_{H^\pm}^{Z \gamma}\, .
\label{eq:10a}
\end{equation}
Defining, as before,
\begin{equation}
c_{H^\pm}^{Z \gamma}
=
\frac{e^2 g}{m_W} \frac{1}{16\pi^2} X_{H^\pm}^{Z \gamma}\, ,
\label{eq:13}
\end{equation}
we get
\ba
X_{H^\pm}^{Z \gamma}
&=&
- \frac{1}{\tan\theta_W}
\frac{\lambda v^2 (1- \tan^2\theta_W)}{m_h^2-m_Z^2}
\left[ \frac{m_Z^2}{m_h^2-m_Z^2}
\left(\vb{12} B_0(m_h^2,m_\pm^2,m_\pm^2)-
B_0(m_Z^2,m_\pm^2,m_\pm^2)
\right)\right.
\nonumber\\[+2mm]
&&
\left.
\hskip 35mm \vb{13}
+
\left(
\vb{12} 2 m_\pm^2 C_0(m_Z^2,0,m_h^2,,m_\pm^2,m_\pm^2,m_\pm^2) + 1
\right)
\right]\, .
\label{eq:11}
\ea
These results agree with Ref.~\cite{HHG,djouadi1},
except for an irrelevant global sign.
See Section~\ref{CompareHHG} for details.

\subsection{Renormalization and gauge invariance}

It is known that a counterterm in needed in order to get a
finite result for this process \cite{Barroso:1985et}.
This happens despite the fact that
there is no tree level coupling $hZ\gamma$.
But,
as explained in ref.~\cite{Barroso:1985et},
the existence of the coupling $hZZ$ and
the renormalization of the mixing $Z\gamma$ leads to a counterterm.
In that work,
the authors were mainly concerned with the divergent part
and did not write the full counterterm.
With our conventions here~\footnote{Our Feynman rules differ from
ref.~\cite{Barroso:1985et}, see ref.~\cite{Romao:2012pq},
and there is a global sign difference.}
we should write instead of
their eq.~(2.16):
\begin{equation}
T^{\nu\nu}_{Z\gamma}
=
- 2 \frac{e g^2 \cos\theta_W M_W}{16 \pi^2}(1+\tan^2\theta_W )\,
g^{\mu\nu} B_0(0,M_W^2,M_W^2)\, ,
\label{eq:24}
\end{equation}
where
\begin{equation}
B_0(0,M_W^2,M_W^2) = \Delta_\epsilon - \ln\frac{M_W^2}{\mu^2},
\quad \Delta_\epsilon = \frac{2}{\epsilon} -\gamma + \ln 4\pi\, ,
\label{eq:25}
\end{equation}
$\gamma$ is the Euler constant, and $\mu$ is the parameter introduced
in dimensional regularization to correct for the fact that the
electric charge is no longer dimensionless in $d\not= 4$.
Apart from a global minus sign,
the divergent part is precisely equal to eq.~(2.16)
of ref.~\cite{Barroso:1985et}.
But there is an important point here concerning the finite
parts.
If we do not take the counterterm as in eq.~\eqref{eq:24},
we will not be able to cancel the dependence on the scale $\mu$
when we sum all the irreducible diagrams.
We have checked this by evaluating all the reducible diagrams and
showing that these sum to the counterterm, that is
\begin{equation}
\sum_{\rm reducible}
=
- 2 \frac{e g^2 \cos\theta_W M_W}{16 \pi^2}
(1 +\tan^2\theta_W ) g^{\mu\nu}
B_0(0,M_W^2,M_W^2)
\equiv T^{\nu\nu}_{Z\gamma} .
\label{eq:26}
\end{equation}
So, in the end, we get a finite result that does not depend on the
scale $\mu$.

\begin{figure}[tbp]
\centering 
\includegraphics[width=.80\textwidth,angle=0]{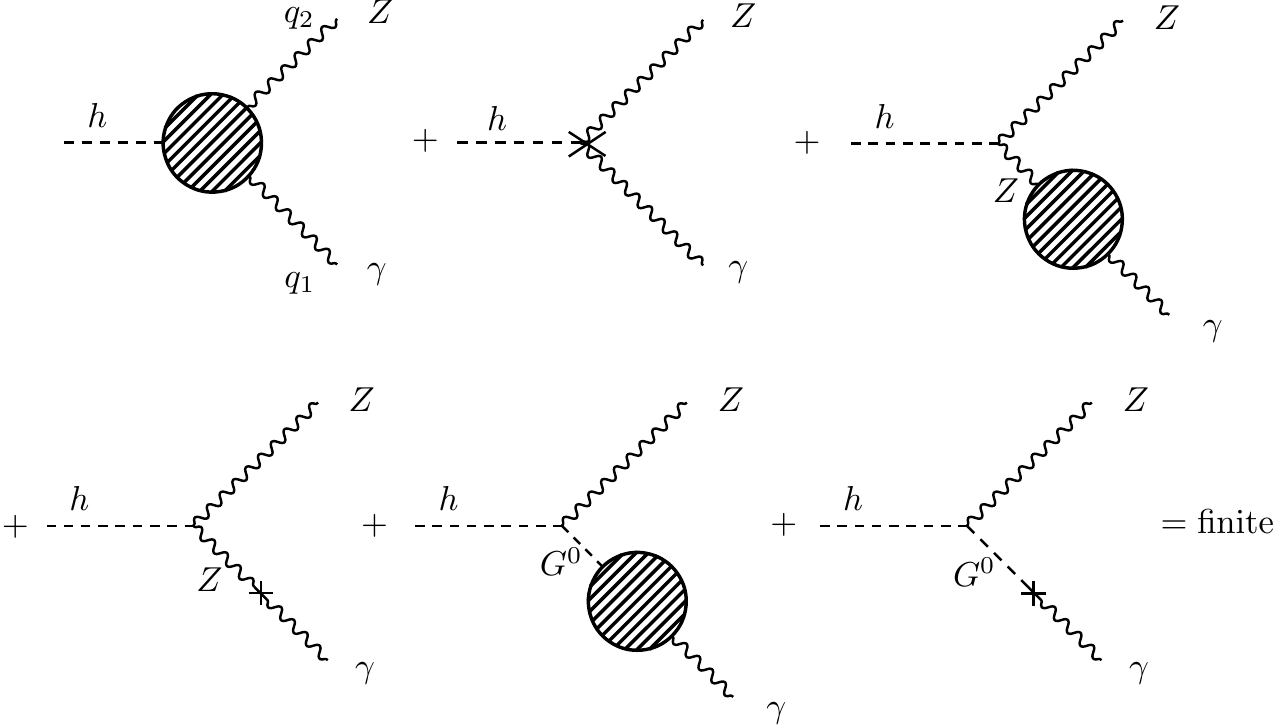}
\caption{\label{fig:1} Classes of one loop diagrams
contributing to $h \ra Z \gamma$.}
\end{figure}

Sometimes it is stated that to get the correct finite result for the
on-shell $hZ\gamma$ three-point function all we have to do is to add
to the irreducible diagrams the sum of the reducible diagrams,
ignoring the counterterms.
For completeness,
we include here an explanation of this statement.
To be precise,
one should add all relevant one loop diagrams,
including reducible, irreducible and counterterms,
as shown in fig.~\ref{fig:1}.

\begin{figure}[tbp]
\centering 
\includegraphics[width=.55\textwidth,angle=0]{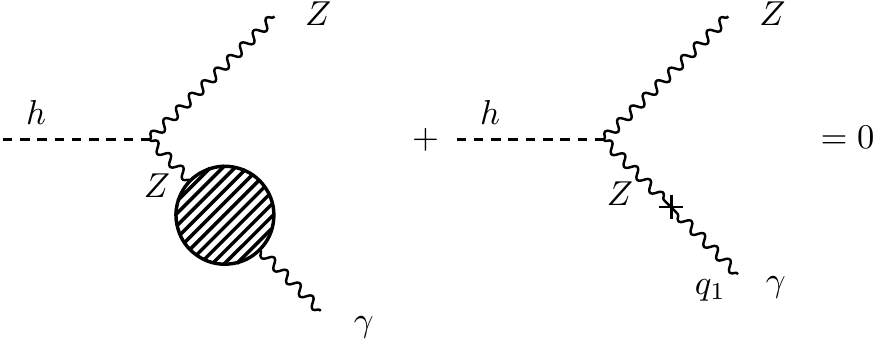}
\caption{\label{fig:2} Sum of diagrams which vanish for a photon
on mass shell.}
\end{figure}

The last two diagrams in fig.~\ref{fig:1},
which involve the Goldstone boson $G^0$,
vanish.
One may keep either of them in or exclude it at will.
Moreover,
the fact that we are using the on mass shell renormalization,
means that the third and fourth diagrams in fig.~\ref{fig:1}
add to zero, as shown diagrammatically in fig.~\ref{fig:2}.

Thus,
we are left with the first two diagrams in fig.~\ref{fig:1}.
We will now show that adding the first and third diagram
in fig.~\ref{fig:1} yields the same result
(as explained above, the fifth and sixth diagrams vanish and,
thus, are optional).
To understand this,
we have to realize that the counterterm
$\delta Z_{hZ\gamma}$ on the second diagram of
fig.~\ref{fig:1} and the counterterm
$\delta Z_{Z\gamma}$ in the photon leg on the fourth diagram of
fig.~\ref{fig:1} are
related.
To show this,
we start with the relevant part of the Lagrangian
\begin{equation}
\mathcal{L}
=
\frac{1}{8} \left( v^2 + 2 v h + h^2\right)
\left[
g^2 W^3_\mu  W_\mu^{3\mu} +
g'^2 B_\mu B^\mu -2 g g' W_\mu^3 B^\mu
\right] + \cdots
 \label{eq:27}
\end{equation}
and perform the shifts
\begin{equation}
g \ra g + \delta g, \quad g'\ra g'+\delta g'\, .
\label{eq:28}
\end{equation}
After using $g'=g \tan\theta_W$ and
\begin{equation}
W^3_\mu =  Z_\mu \cos\theta_W + A_\mu \sin\theta_W,
\quad
B_\mu = -Z_\mu \sin\theta_W + A_\mu \cos\theta_W,
\label{eq:29}
\end{equation}
we get
\ba
&&
\left[g^2   W^3_\mu  W_\mu^{3\mu}\right.
\left.
+ g'^2 B_\mu B^\mu -2 g g' W_\mu^3 B^\mu
\right]
\nonumber\\
&&
\ \ \
\ra
\frac{g^2}{\cos^2\theta_W} Z_\mu Z^\mu + 2g Z_\mu Z^\mu (\delta g +
\delta g' \tan^2\theta_W) + 2 g Z_\mu A^\mu (\delta g \tan\theta_W -
\delta g').
\label{eq:30}
\ea
As the mixing term in $Z_\mu A^\mu$ is already
first order in the corrections,
we do not need to perform the shifts
in $v$ and $h$ to get,
finally,
\begin{equation}
\delta Z_{Z\gamma} = \frac{1}{2}\, v\, \delta Z_{hZ\gamma}.
\label{eq:31}
\end{equation}

\begin{figure}[tbp]
\centering 
\includegraphics[width=.55\textwidth,angle=0]{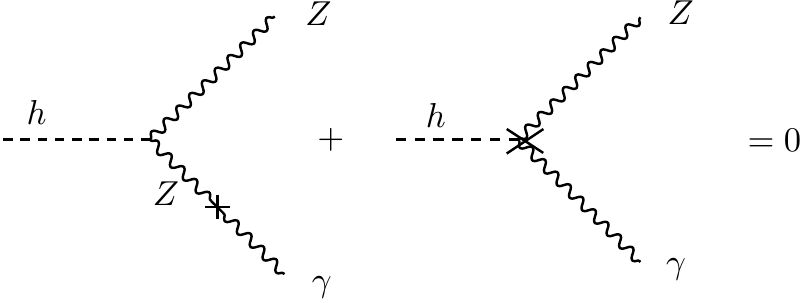}
\caption{\label{fig:3} Diagrammatic form of evaluating the
fourth diagram in fig.~\ref{fig:1}
via eq.~\eqref{rel_counter}.}
\end{figure}

Let us now evaluate the diagram with the counterterm in
fig.~\ref{fig:2}.
We have,
for on-shell photon ($q_1^2=0$),
\be
\ i \frac{g}{\cos\theta_W} M_Z
\frac{-i}{-M_Z^2} \ i\, \delta Z_{Z\gamma}\,
=\,
-i\, \frac{g}{M_W} \,\delta Z_{Z\gamma}
=
-i \delta Z_{hZ\gamma},
\label{rel_counter}
\ee
where we have used eq.~\eqref{eq:31} and
$M_W= \frac{1}{2}\, g\, v$.
We obtain the result in fig.~\ref{fig:3}.

\begin{figure}[tbp]
\centering 
\includegraphics[width=.55\textwidth,angle=0]{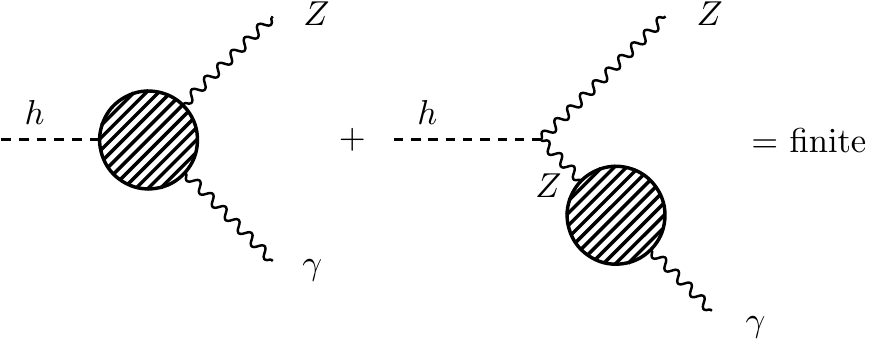}
\caption{\label{fig:4} Diagrammatic form of evaluating the
fourth diagram in fig.~\ref{fig:1}
via eq.~\eqref{rel_counter}.}
\end{figure}

Having established that the calculation can be performed
exclusively with the first and second diagrams in fig.~\ref{fig:1},
and combining figs.~\ref{fig:2} and \ref{fig:3},
we obtain the result in fig.~\ref{fig:4},
which we were seeking.
That is:
as often stated,
one can add all reducible and irreducible diagrams,
ignoring the counterterms.

\section{Widths for loop decays}

The total width is given by
\begin{equation}
  \label{eq:5}
  \Gamma = \frac{1}{8\pi} \frac{|\vec q_1|}{m_h^2} \overline{|M|^2}\, .
\end{equation}

\subsection{\texorpdfstring{$h \ra \gamma \gamma$}{h --> gamma gamma}}

In this case, $|\vec q_1|= m_h/2$, and
\ba
\overline{|M|^2}
&=&
\left( \frac{e g}{16\pi^2 m_W}\right)^2
\left[
    |X_F^{\gamma\gamma}+ X_W^{\gamma\gamma} + X_H^{\gamma\gamma}|^2
      \left( q_1\cdot q_1 g_{\mu\nu}-q_{1\mu} q_{2\nu}\right)
      \left( q_1\cdot q_1 g_{\mu'\nu'}-q_{1\mu'} q_{2\nu'}\right)
\right.
\nonumber\\[+2mm]
&&
\hspace{30mm}
      (-g^{\mu\mu'})(-g^{\nu\nu'})
\nonumber\\[+2mm]
&&
\left.
\hspace{25mm}
 + |Y_F^{\gamma\gamma}|^2\ \epsilon_{\mu\nu\alpha\beta} q_1^\mu q_2^\nu
 \epsilon_{\mu'\nu'\alpha'\beta'} q_1^{\mu'} q_2^{\nu'}
 (-g^{\alpha\alpha'})(-g^{\beta\beta'})
\right]
\nonumber\\[+2mm]
&=&
\left( \frac{e g}{16\pi^2 m_W}\right)^2 \frac{m_h^4}{2}
\left(
  |X_F^{\gamma\gamma} + X_W^{\gamma\gamma} + X_H^{\gamma\gamma}|^2 +
  |Y_F^{\gamma\gamma}|^2
\right)\, .
\label{eq:6}
\ea
Putting everything together, and including the factor 1/2 for identical
particles, we get the final result
\begin{equation}
\Gamma(h \ra \gamma\gamma) = \frac{G_F\alpha^2 m_h^3}{128\sqrt{2}\pi^3}
\sum_f \left( |X_F^{\gamma\gamma}+ X_W^{\gamma\gamma} +
X_H^{\gamma\gamma}|^2 + |Y_F^{\gamma\gamma}|^2 \right)\, .
\label{eq:7}
\end{equation}

\subsection{\texorpdfstring{$h \ra Z \gamma$}{h --> Z gamma}}
\label{app:hZgamma_widths}

Now, we have $|\vec q_1|= (m_h^2-m_Z^2)/(2 m_h)$,
and
\ba
\overline{|M|^2}
&=&
\left( \frac{e g}{16\pi^2m_W}\right)^2
\left[
    |X_F^{Z \gamma}+ X_W^{Z \gamma} + X_H^{Z \gamma}|^2
      \left( q_1\cdot q_1 g_{\mu\nu}-q_{1\mu} q_{2\nu}\right)
      \left( q_1\cdot q_1 g_{\mu'\nu'}-q_{1\mu'}
        q_{2\nu'}\right)
\right.
\nonumber\\[+2mm]
&&
\hspace{30mm}
(-g^{\mu\mu'})
\left(-g^{\nu\nu'} + \frac{q_2^\nu q_2^{\nu'}}{m_Z^2}
\right)
\nonumber\\[+2mm]
&&
\left.
\hspace{25mm}
 + |Y_F^{Z \gamma}|^2 \epsilon_{\mu\nu\alpha\beta} q_1^\mu q_2^\nu
 \epsilon_{\mu'\nu'\alpha'\beta'} q_1^{\mu'} q_2^{\nu'}
 (-g^{\alpha\alpha'})
\left(
-g^{\beta\beta'} +
 \frac{q_2^\beta q_2^{\beta'}}{m_Z^2}
\right)
\right]
\nonumber\\[+2mm]
&=&
\left( \frac{e g}{16\pi^2 m_W}\right)^2 \frac{(m_h^2-m_Z^2)^2}{2}
\left(
  |X_F^{Z \gamma}+ X_W^{Z \gamma} + X_H^{Z \gamma}|^2 +
  |Y_F^{Z \gamma}|^2
\right)\, ,
\label{eq:6a}
\ea
and, for the final width,
\begin{equation}
\Gamma(h \ra Z \gamma) =
\frac{G_F \alpha^2 m_h^3}{64\sqrt{2}\pi^3}
 \left(1 - \frac{m_Z^2}{m_h^2}\right)^3
\sum_f \left( |X_F^{Z \gamma}+ X_W^{Z \gamma}
+ X_H^{Z \gamma}|^2 + |Y_F^{Z \gamma}|^2 \right)\, .
\label{eq:7b}
\end{equation}

\section{Relation between the Passarion-Veltman functions
and loop functions}
\label{app:relation}

When we compute the one-loop diagrams, as we did, using
\texttt{FeynCalc} \cite{feyncalc:1992}, the result is naturally
presented in terms of the well-known Passarino-Veltman
integrals \cite{passarino:1979jh}. These are in general complicated
functions of the external momenta and masses and usually only possible
to be expressed in terms of very complicated functions. Normally it is
better to evaluate them numerically and for that there is the package
\texttt{LoopTools} \cite{looptools:2001}. However for special
situations, like zero external momenta or equal masses in the loops,
these loop integrals have simpler forms and can be expressed in terms
of simple functions. This is the case for the loops studied here and
we present in this appendix the relations of these Passarino-Veltman
integrals with other representations found in the literature.

\subsection{The integrals for
\texorpdfstring{$h \ra \gamma \gamma$}{h --> gamma gamma}
}

In this decay,
all results can be expressed in terms of the Passarino-Veltman
integral $C_0(0,0,m_h^2,m^2,m^2,m^2)$,
where $m$ is the mass of the particle running in the loop.
We have already given in eq.~\eqref{eq:4} the relation
with the function $f(\tau)$ defined in
the Higgs Hunter's Guide \cite{HHG},
\begin{equation}
  \label{eq:4a}
  C_0(0,0,m_h^2, m^2,m^2,m^2) = - \frac{\tau f(\tau)}{2 m^2}, \quad
  \tau=\frac{4 m^2}{m_h^2},
\end{equation}
where $f(\tau)$ is defined in eq.~\eqref{eq:9}.

\subsection{The integrals for
\texorpdfstring{$h \ra Z \gamma$}{h --> Z gamma}
}
\label{CompareHHG}

In the Higgs Hunter's Guide \cite{HHG},
a different set of integrals, $I_1(a,b)$ and $I_2(a,b)$ were
introduced. They are defined as follows:
\ba
I_1(a,b)
&=&
\frac{a b}{2(a-b)} + \frac{a^2
  b^2}{2(a-b)^2}\left[\vb{11} f(a)-f(b) \right] + \frac{a^2
  b}{(a-b)^2}
\left[ \vb{11} g(a)-g(b) \right]\, ,
\\[+2mm]
I_2(a,b)
&=&
-\frac{a b}{2(a-b)} \left[\vb{11} f(a)-f(b) \right]\, ,
\label{eq:14}
\ea
where $f(\tau)$ was defined in eq.~\eqref{eq:9},
and $g(\tau)$ is given by
\begin{equation}
g(\tau)= \left\{
    \begin{array}{ll}
      \sqrt{\tau -1}\
      \sin^{-1} \left(\sqrt{1/\tau}\right), &
      \ \ \text{if}\
      \tau\ge 1\ ,\\[+2mm]
      \frac{1}{2} \sqrt{1-\tau}\ \left[ \displaystyle
        \ln\left(\frac{1+\sqrt{1-\tau}}{1-\sqrt{1-\tau}}\right)
        -i \pi \right] ,
        &
      \ \ \text{if}\ \tau<1\, .
    \end{array}
\right.
\label{eq:9b}
\end{equation}
Comparing their results with our results and those of
ref.~\cite{Barroso:1985et},
we get
\begin{equation}
I_1(\tau,\lambda) = 4 J_2(\beta_Z,\beta_H)\\[+2mm], \quad
I_2(\tau,\lambda) =  J_1(\beta_Z,\beta_H)\, ,
\label{eq:20}
\end{equation}
where
\begin{equation}
\tau
= \frac{4 m^2}{m_h^2},
\quad \lambda = \frac{4 m^2}{m_Z^2},
\quad \beta_Z= \frac{m_Z^2}{m^2},
\quad \beta_H= \frac{m_H^2}{m^2}\, ,
\label{eq:21}
\end{equation}
and $m$ is any mass running in the loops. Again, we have numerically
checked that these relations hold for any value of the arguments.

To compare our results in terms of the Passarino-Veltman functions
with those of ref.~\cite{HHG},
we notice that
\ba
&&
C_0(m_Z^2,0,m_h^2,m^2,m^2,m^2)
=
- \frac{1}{m^2}
I_2(\tau,\lambda)\, ,
\\[+2mm]
&&
\Delta B_0
=
-\frac{m_h^2-m_Z^2}{m_Z^2}
- \frac{(m_h^2-m_Z^2)^2}{2 m^2 m_Z^2}\, I_1(\tau,\lambda)
+ 2\, \frac{m_h^2-m_Z^2}{m_Z^2}\, I_2(\tau,\lambda)\, .
\label{eq:23}
\ea
We have checked these equations numerically with the help of the package
\texttt{LoopTools}~\cite{looptools:2001}

Using these relations, one can check that our eqs.~\eqref{eq:15a},
\eqref{eq:16} and \eqref{eq:11} agree with eqs. (C.12), (C.13) and
(C.14) of ref.~\cite{HHG} up to an overall sign.
We notice that
our coupling to the charged Higgs translate into their notation
\begin{equation}
\lambda v \ra - R^h_{H^\pm} \, .
\label{eq:15}
\end{equation}
There is no equivalent result to our eq.~\eqref{eq:15b} in
Ref.~\cite{HHG}, but we are in agreement with
Ref.~\cite{djouadi1} up to global signs.
However we warn the
reader that the definitions of $I_1$, $I_2$ and $g(\tau)$ in
eqs. (2.55) and (2.56) of Ref.~\cite{djouadi1} are not consistent.

\section{Production and decay involving gluons}

Relating with the expression for the $\gamma\gamma$ decay,
we find
\begin{equation}
\Gamma(h \ra gg) = \frac{G_F\alpha_S^2 m_h^3}{64\sqrt{2}\pi^3}
\left( |X_F^{gg}|^2 + |Y_F^{gg}|^2 \right)\, ,
\end{equation}
where
\ba
X_F^{gg}
&=&
- \sum_q 2 a_q\, \tau_q \left[ 1 + (1-\tau_q)
f(\tau_q)\right]\, ,
\nonumber\\[+2mm]
Y_F^{gg}
&=&
-\sum_q  2 b_q\, \tau_q f(\tau_q)\, ,
\ea
and the sums run only over quarks $q$.

Similarly,
\be
\sigma (g g \rightarrow h)
=
\frac{G_\mu \alpha_s^2}{512 \sqrt{2} \pi}
\left( |X_F^{gg}|^2 + |Y_F^{gg}|^2 \right)\, .
\ee
These are dominated by the triangle with top quark in the loop,
and, depending on $\tan{\beta}$,
also by the triangle with bottom quark in the loop.
Thus,
we can use
\be
\frac{\sigma (g g \rightarrow h)}{\sigma^{\textrm{SM}} (g g \rightarrow h)}
=
\frac{|a_t\, A_{1/2} (\tau_t) + a_b\, A_{1/2} (\tau_b)|^2 +
|b_t\, A^A_{1/2} (\tau_t) + b_b\, A^A_{1/2} (\tau_b)|^2 }{
|A_{1/2} (\tau_t) + A_{1/2} (\tau_b)|^2}\, ,
\ee
where
\ba
A_{1/2} (\tau_q) =
&=&
2 \tau_q \left[ 1 + (1-\tau_q) f(\tau_q)\right]\, ,
\nonumber\\
A_{1/2}^A(\tau_q)
&=&
2 \tau_q f(\tau_q)\, .
\ea

\acknowledgments

We are grateful to Rui Santos for many discussions related to the
Higgs production channels and to Augusto Barroso
for discussions on the renormalization of the
$h Z \gamma$ vertex.
This work was partially supported by FCT - \textit{Funda\c{c}\~{a}o para a
Ci\^{e}ncia e a Tecnologia}, under the projects
PEst-OE/FIS/UI0777/2013 and CERN/FP/123580/2011.  D.~F. is also
supported by FCT under the project EXPL/FIS-NUC/0460/2013.



\end{document}